\pdfoutput=1 
\documentclass{article}

\usepackage{amsmath,amssymb,amsfonts}

\usepackage{graphicx}

\usepackage{listings}

\usepackage[usenames]{color}  % allows colored text

\newtheorem{rule-of-thumb}[theorem]{Definition} % Definite {rule-of-thumb}

\begin{document}
\title{Meteorology and Oceanography on a Flat Earth}

\author{John P. Boyd \\
Department of Climate \& Space
Sciences and Engineering  \\University of
Michigan, 2455 Hayward Avenue, Ann Arbor MI 48109 \\
jpboyd@umich.edu}

\maketitle

\begin{abstract}
To build insight into the atmosphere and ocean, it is useful to apply qualitative reasoning to 
predict the geophysical fluid dynamics of worlds radically different from our own such as exoplanets, earth in Nuclear Winter, other solar system worlds, and far future terrestial climates. Here,  we look
at atmospheric and oceanic dynamics on a flat earth, that is a disc-shaped planet  rather like Sir Terry Pratchet's fantasy Discworld. Altough this has the disadvantage that this geometry is completely imaginary, there is a rich 
array of videos by flat earth evangelists whose errors illuminate how concepts can be misconceived and misapplied by amateurs and 
freshman science studients. As such, this case is very useful to geophysics instructors.
   We show that weather and ocean flows on a flat, nonrotating earth and a rotating spherical planet are wildly different. These differences are 
a crushing debunk of the flat earh heresy, if one were needed. The ``high contrast" of these very different atmospheres and oceans is valuable in
instilling the open-mindedness that is essential to understand excoplanets, Nuclear Winter and Post-Climate-Apocalypse earth.
\end{abstract}

\tableofcontents

\bigskip
\bigskip

Dedicated to William R. Kuhn, Prof. Emeritus, University of Michigan, for more than forty years as colleague 
and exceptional teacher.

\bigskip

\begin{quotation}
 ``The decay time of the atmosphere [after the sun was turned off] was found to be considerably longer than previous estimates in the literature because the dissipation rates declined proportionately faster than the energy.
\end{quotation} \hspace*{0.5 in} ------  B. G. Hunt in the abstract of his 1976 article, ``On the Death of the Atmosphere" \cite{Hunt76}
\bigskip

\section{Introduction}

\subsection{Radical meteorology}

It is said that one who only knows a single language knows not his own. Much research 
on exoplanets and on the planets in our own solar system, other than earth, is motivated
   by a similar philosophy.

A cosmology in which the earth is a flat, nonrotating disk of finite size is irreconcilable with 
 mainstream physics. Such a disk world would rapidly and irreversibly collapse into a sphere. 
   Nevertheless, working out the atmospheric and oceanic dynamics of such a world is useful because 
it illuminates the tremendous importance of the earth's rotation to weather on the spherical
planet whose surface we actually inhabit.

Exoplanets and the continuing exploration of the  solar system are challenging atmospheric scientists to think far outside the box. This article is in that spirit. However, there is a long tradition of ``radical meteorology" that far preceded exoplanets. 
 My colleague, Bill Kuhn, was fond of asking doctoral students on qualifying exams very open-ended questions:  Describe the atmosphere of Titan under early `Dim Sun' " conditions? What happens to Mars if the 
sun dies? These questions tested the students' ability to syncretize a grab bag of dynamics, chemistry, 
radiation and instrumentation into a comprehensive view of the subject. 

   ``Radical meteorology" has its uses on the research frontier as well, even for earth.  Using a hemispheric
general circulation model, B. G. Hunt showed showed that even    fifty days after the sudden termination of all
 incoming solar radiation, a quarter of the initial kinetic energy  remained in the atmosphere. 
 Similar though slightly less drastic reductions in solar output must be examined to study 
asteroid extinctions \cite{Reid81}, volcanic eruptions \cite{Hunt77} and ``Nuclear Winter", which is the decrease in insolation due to smoke and 
atmospheric soot generated by thermonuclear war \cite{Crutzen84}. The earth's rotation rate is 
also fair game for climate studies such as \cite{KuhnWalkerMarshall89,Hunt79,Simmonds85}.
 In this tradition, we analyze flow on a disk world.

The recent article by de Marez and Le Corre \cite{MarezLeCorre20} compares numerical simulations 
of the North Atlantic Gyre on both flat and spherical earths. They show that the spherical results 
are in good agreement with drift observations, discussed further below. The flat earth results 
 are quite different form the observed ocean even though the authors assumed that the 
disk was rotating. Rotation brings the flat earth model closer to reality, but a rotating disk is 
anathema to flat earthers because an underlying tenet is: ``Trust your own senses rather 
than mainstream science". If the disk is spinning, then the human senses have been fooled, 
and the notion that the earth must be flat because it appears so to the nakd eye is
  cast in doubt.

\subsection{Impossibilities and ``hard" science fiction}

\begin{quotation}
Writing a science fiction story is fun, not work. ...the fun...lies in treating the whole thing as a game.... [T]he rules must be quite simple. For the reader of a science-fiction story, they consist of finding as many as possible of the author's statements or implications which conflict with the facts as science currently understands them. For the author, the rule is to make as few such slips as he possibly can... Certain exceptions are made [e.g., to allow travel faster than the speed of light], but fair play demands that all such matters be mentioned as early as possible in the story...
\end{quotation}
 \hspace*{0.5 in} ------- Hal Clement (Harry Clement Stubbs (1922-2003) in the nonfiction essay 
``Whirligig World" included in most editions of his novel \emph{Mission of Gravity} \cite{Clement54}.

The student must approach conceptual exercises such as weather-on-a-flat-earth in the spirit of 
so-called ``hard" science fiction. What distinguishes ``hard" science fiction from fantasy and 
soft science fiction is the determinantion, expressed by Clement, of working out the science 
as carefully as possible while allowing one or more violations of contemporary science for 
the sake of the story. Faster-than-light (FTL) travel is the most common exception; hard sf is consoled
by the thought that FTL may be possible in a few centuries.\footnote{Clement's planet Mesklin has a day 
only seventeen minutes long, an oblate spheroid more a disk than a sphere. I myself have explored 
a toroidal planet in \cite{BoydMoonbow,BoydEarthflight}, an airship-friendly world in
   \cite{BoydHero,BoydVictory,BoydTempest} and starless planets (``anasters") in \cite{BoydElectricSouls,BoydIcosahedral}.}

No such comfort is available for the impossibilities of flat earth as explained below. Outsider science 
is sometimes an intriguing venue for science fiction, ocasionally good art, but almost never good science. We can neverthess proceed to
analyze geophysical  fluid flow on a disc in the spirit of Clement and other hard sf writers: to stretch 
our minds.

  \section{Gravity}

A crucial assumption of all flat earth theories is that gravity operates on the disk as a force perpendicular to the 
surface of the disk, accelerating all objects downward towards the solid surface of the disk at 9.81 
meters/second$^{2}$. 

The gravity force vector should of course be directed toward the 
center of mass of the planet, not normal to a disk. Except at the center of the disk, gravity   will generate a component
  which is radially inward towards the ``North Pole" of the disk. We must assume an imaginary 
force, ``fairy-g", which is everywhere downward on the disk, in order to make useful comparisons 
and analogies with the flow on a rotating planet. (See Appendix C for further discussion.)

\section{Flat Earth Geometry \& Cartography}

\subsection{Flat earth maps}

Flat earth theorists have proposed a number of geometries including an infinite plane, Derren Nesbit's 
square with boundary conditions ( periodicity in longitude, glide reflection at the poles) that make the surface topologically equivalent to the sphere, a square repeated periodically over the infinite plane, various species of sphere projected to a disk by means of a bipolar map  projection and a ``Supernatural Construct" where the last is a frank admission that no ordinary or even extraordinary flat surface is free 
of contradictions. By far the most common flat earth concept is a flat disk of finite diameter. \footnote{This is surmounted by a dome 
which supports the stars and planets. A planet-sized half-sphere would collapse unless made of a material 
several orders of magnitude stronger than any earthly material.  Fortunately this dome of, let us call it  
``unobtainabletransparentarium" or ``fairy-glass", is irrelevant to tropospheric meteorology.}
 This is almost universally depicted by mapping continents to the disk using the aximuthal 
equidistant projection, known in flat earth circles as the ``Gleason map" \cite{Gleason93}.

The disk is always assumed to be {motionless}.An important justification for the flat earth concept is that human senses are completely trustworthy; therefore, a planet that is percieved as flat and motionless must be 
flat and motionless.

\begin{figure}[b]
\centerline{\includegraphics[scale=0.5]{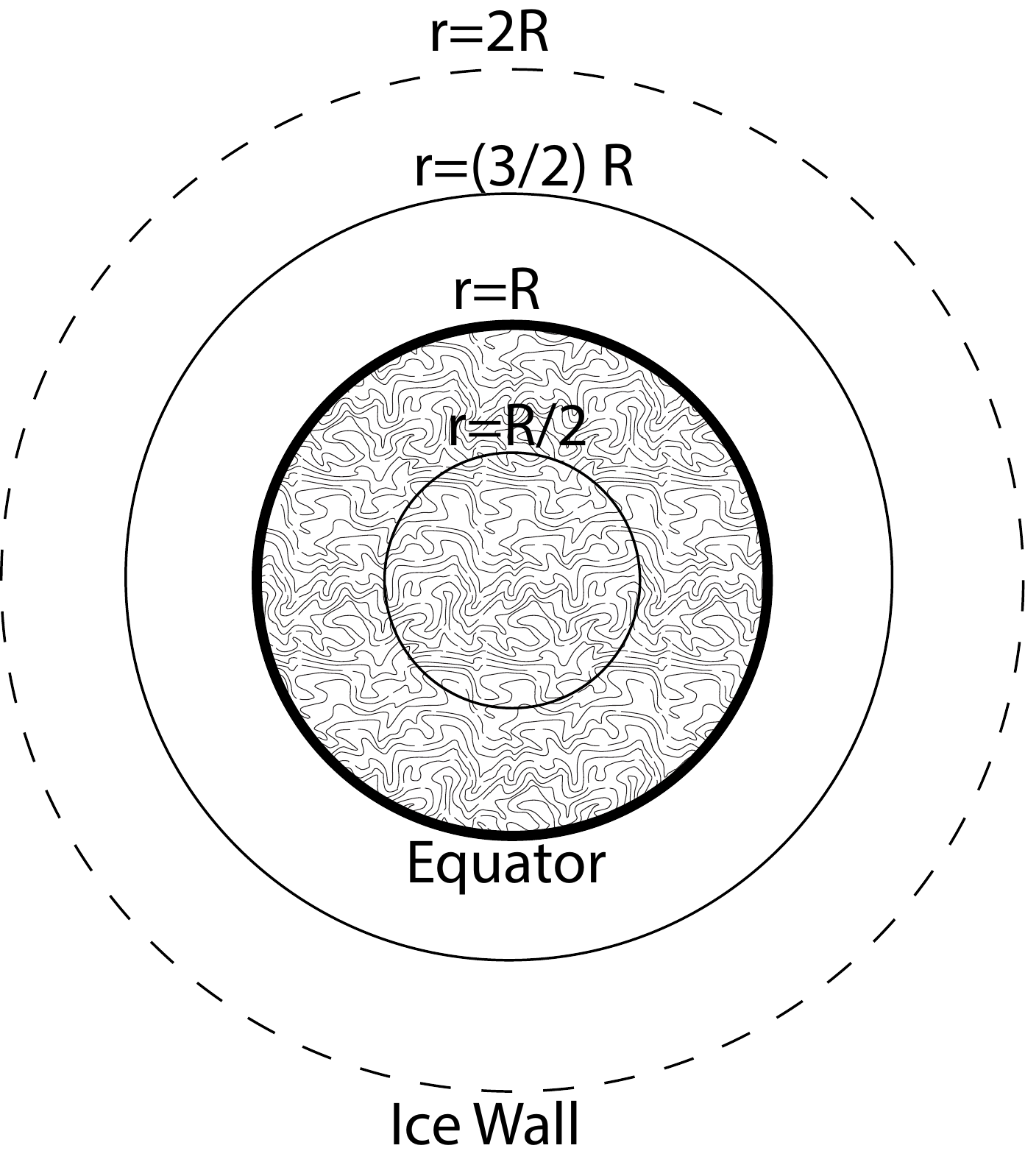}}
\caption{The geometry of the disk of the Flat Earth. The ``disk Northern Hemisphere" is shaded. 
The North Pole is at the center of the diagram; in terms of polar coordinates $(r, \theta)$, this point is at 
$r=0$. The equator is the thick solid curve, a circle of radius $R$ where $R$ is the circumference of the 
earth. The dashed circle at $r=2R$ is the South Pole, which is stretched into a circular ``ice wall". 
The thin circles of radius $r=R/2$ and $r=(3/2) R$ represent the latitude circles with latitude = 45 degrees N. and 45 degrees S., respectively. The ``disk Southern Hemisphere" is the unshaded annulus; note that the 
area of the ``Southern Hemisphere" is three times larger than that of the ``Northern Hemisphere". The latitude circle at 60 degrees south latitude on the disk has a circumference of $(5/3) E$ where $E$ is the equatorial circumference but the same circle on the sphere has a circumference of $E/2$, a factor of $3.33$ difference.}
\label{FigFlatEarth_Sch_Map}
% Drawn by me in Illustrator as FlatEarth_Sch_Map
\end{figure}

Disk cartography uses polar coordinates $(r, \theta)$ with the radial coordinate $r$ as the distance to the North Pole and $\theta$ as the polar angle. The latter is a cyclic coordinate with $\theta \in [0, 2 \pi]$, identical 
with longitude. The radial coordinate $r$ is analogous to colatitude. It is convenient to change to a 
scaled radial coordinate
\begin{eqnarray}
r = R \rho, \qquad \rho \in [0, 2]
\end{eqnarray}
where $R$ is the distance from the North Pole to the equator.
We shall define the disk that is the direct product of $\rho \in [0, 1]$ and $\theta \in [0, 2 \pi]$ as the ``[disk] Northern Hemisphere" and the annulus which is all points with $\rho \in [1, 2]$ as the ``[disk] Southern Hemisphere" as illustrated in \ref{FigFlatEarth_Sch_Map}.

 % **********************  FIGURE  ***********
% ***************************************

\subsection{Discussion of why it is impossible to simultaneously match the pole-to-equator distance $R$ and the equatorial circumference $E$ on the disk to the corresponding values on the globe}

On a sphere, the pole-to-equator distance (along a meridian) is equal to \emph{one-quarter} of 
the circumference of the equator, 10,000 km and 40,000 km, respectively. If $R$ is the 
distance from the pole to the equator on a disk, then the \emph{disk} equator has a circumference 
of $E=2 \pi R$. This make it impossible to simultaneously match the pole-to-equator distance $R$ and the equatorial circumference $E$ on the disk to the corresponding values on the globe 
as enumerated in Table~\ref{TabDiskRandE} .

If we choose to match the pole-to-equator distances by choosing $R=10,000$ km on the disk,
the disk equator has a circumference of 62,800 km. Such a disk earth has an equatorial circumference 
22,800 km larger than the generally accepted value. Interesting!

If we choose to match the equatorial circumferences instead, then the pole-to-equator distance on the 
disk is shortened to 6366 km. However, in the metric system, the meter was defined to be 1/10,000,000 of the pole-to-equator distance. Did all the world's navigators and pilots, by land and by air, somehow  
not notice that pole-to-equator distance was shorter, by over 3600 km, than the value 
published in the standard navigational and geographic tomes. Interesting!

% *********** TABLE 9-1 *********************************
\begin{table}[h] \caption{\label{TabDiskRandE} The North-Pole-to-equator distance $R$ and the circumference 
of the equator $E$ for three different models}
% {\footnotesize This is how to add a note to table.}
%  \vspace{5pt}
\begin{center} \begin{tabular}{|c|c|c|} \hline
Model & $R$ & $E$ \\ \hline
Sphere & 10,000 km & 40,000 km \\
Small Disk & 6366 km & 40,000 km \\
Large Disk & 10,000 km & 62,800 km \\ \hline
 \end{tabular} \end{center} \vspace{5pt} \end{table}  
% ******************************************************

\begin{figure}[h]
\centerline{\includegraphics[scale=0.7]{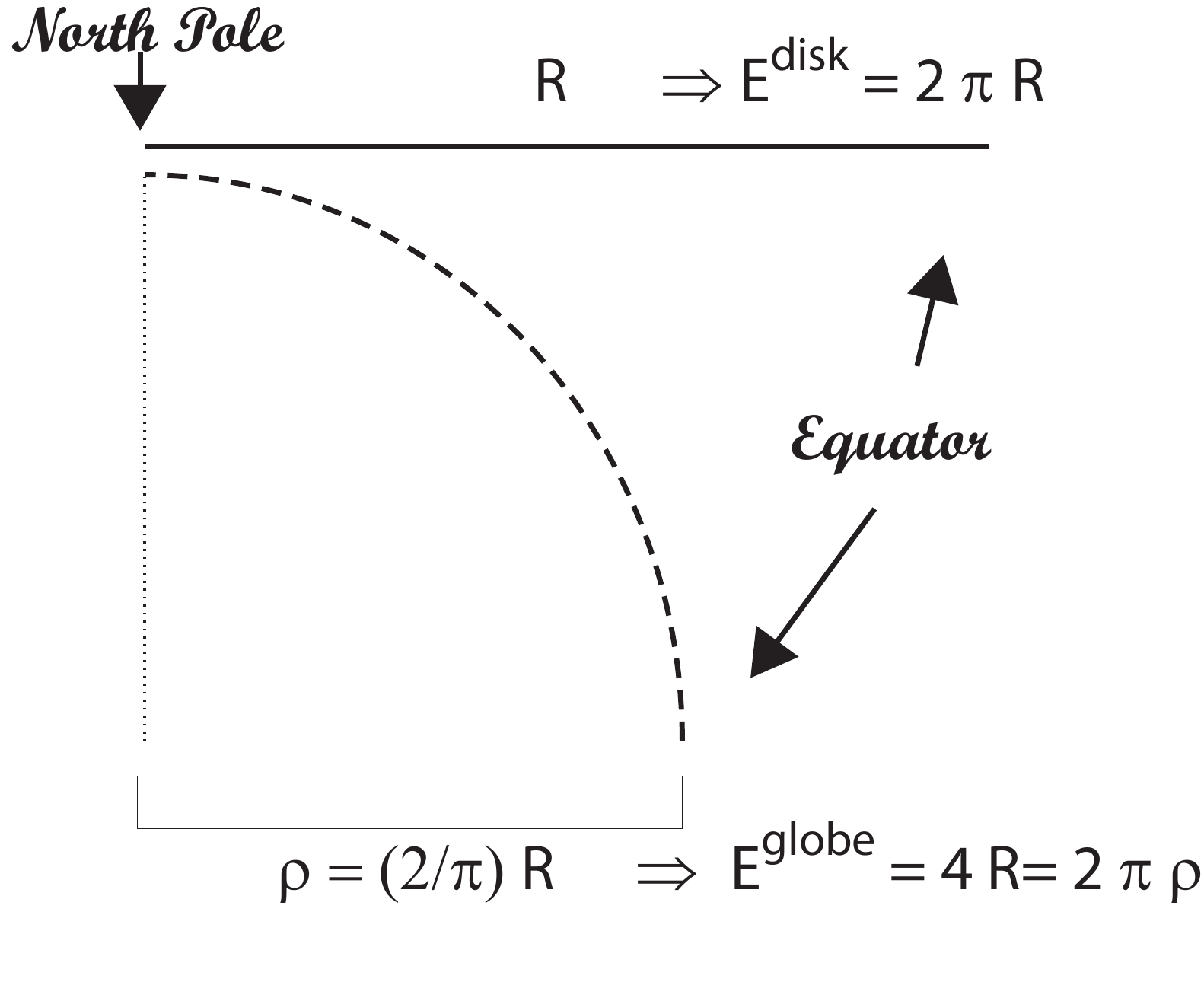}}
\caption{The vertical dotted line is the spin axis of the globe. The straight solid horizontal line of length
$R$ connects the North Pole with the equator on the  flat disk.  The dashed curve connects pole-to-equator on the sphere and 
is also of length $R$. Denote the radius of the globe earth by $\rho$. Because of the curve, the globe equator is only a distance $\rho=(2/\pi) R$ away from the rotation axis. The circumference of 
the equator on the disk is $E^{disk}=2 \pi R$; the equatorial circumference on the globe is$ (2/\pi)$ smaller,
$E^{globe} = 4 R = 2 \pi \rho$.  
}
\label{FigFlat_earth_disk_two_poletoequator_curves}
% Flat_earth_disk_two_poletoequator_curves.m; text added in Illustrator
\end{figure}

The reason  that $E=4 R$ on the sphere but $E=2 \pi R$ on the disk is that the  pole-to-equator  curve on the surface of a 
sphere  is \emph{curved} whereas the ray from pole to equator on the disk is \emph{flat}. 
The equatorial terminus of the curve on the sphere is much closer (by $2/\pi)$)  to the axis of the globe than 
the equatorial end of the line on the disk is to the North Pole.
 Fig.~\ref{FigFlat_earth_disk_two_poletoequator_curves} is a visual comparison.

 % **********************  FIGURE  ***********
% ***************************************

\subsection{Circumference of a circle of latitude}

The isolines of distance to the North Pole are concentric circles on both the sphere and on the disk. 
  We shall extend the label of these isolines from the sphere to the disk, and call these 
``circles of latitude" in either case. The circumference of these circles of latitude is quite 
different for the sphere and the disk:
\begin{eqnarray}
C(latitude) & = & E^{globe} \, \cos(latitude) = 4 R \cos(latitude) \qquad \text{Sphere} \\
     & = & E^{disk}  \left( 1 -  \dfrac{2}{\pi}  latitude \right)
\end{eqnarray}
where latitude is in radians [=360 degrees/($2 \pi$) ].

% **********************  FIGURE  ***********
\begin{figure}[h]
\centerline{\includegraphics[scale=0.7]{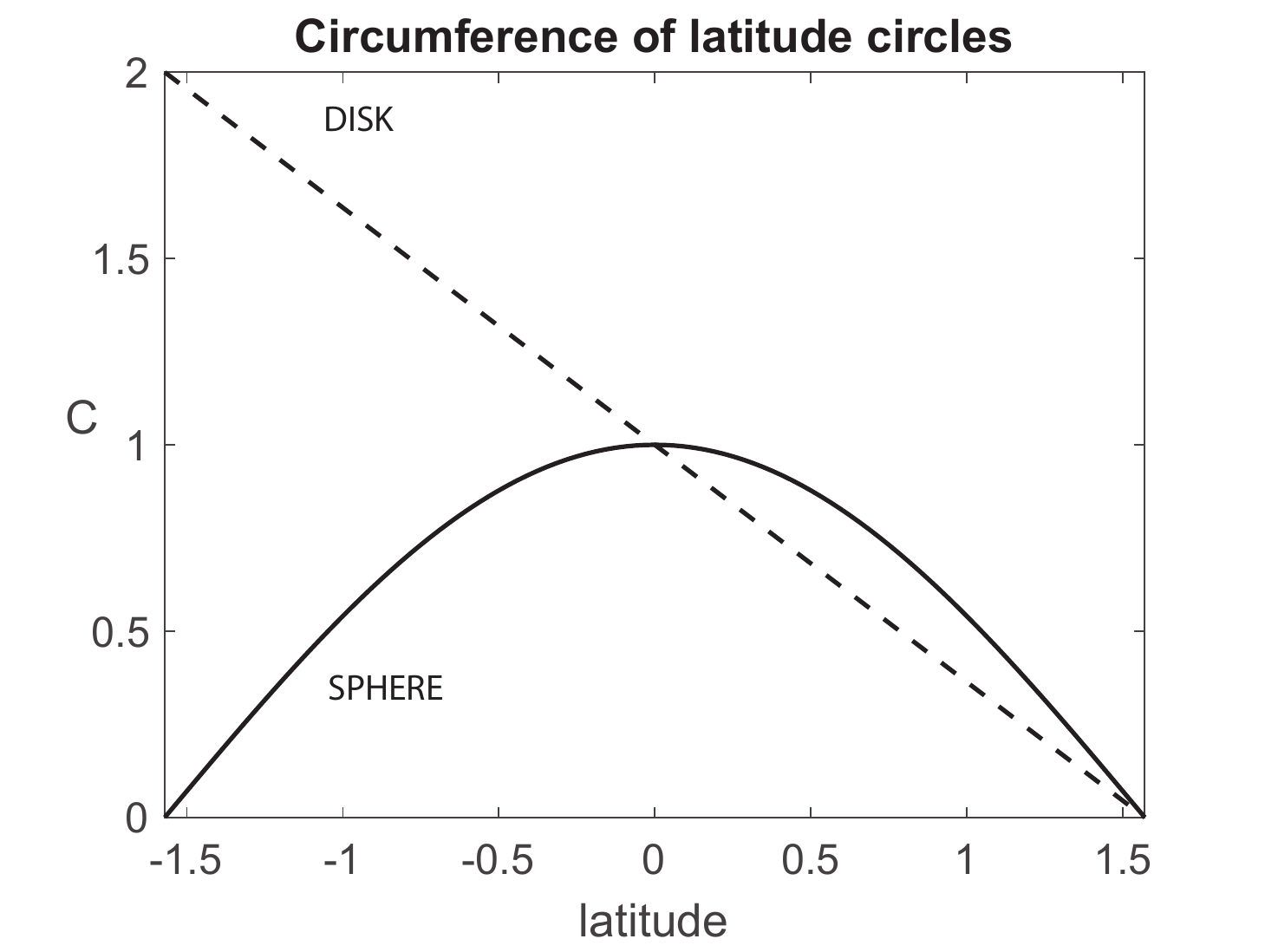}}
\caption{ Solid: circumference of latitude circles as a function
 of latitude in radians. Dashed: same but for the disk. 
The figure assumes that the circumferences of the equators 
are the same.
}
\label{FigCircumference_Circles_Latitude}
% Circumference_Circles_Latitude.m; text added in Illustrator
\end{figure}  
% ***************************************

Fig.~\ref{FigCircumference_Circles_Latitude} shows that the circumference of a latitude circle 
increases monotonically as we progress from the North Pole (right of the diagram) to the 
South Pole (left). In contrast, the circumference of a latitude circle on the globe is a 
 maximum on the equator and diminishes to zero at the South Pole.

The Australian cites of Brisbane (-27.5 degrees S, 153 degrees E, and Perth  ( -32 degrees S, 115.5E) are almost on the same circle of latitude. Splitting the difference by using the circumference of $-30$ degrees, the distance between the two cities is the circumference of that latitude circle multiplied 
by (153 - 115.5)/360. One obtains 5480 km on the disk, but only 3560 km on the globe. The 
actual distance is 3600 km.

The disk thus gives the Perth-to-Brisbance by a number 49 \% too large.  Presumably it is a matter 
of intense interest to airlines and long-haul truckersthat the distance beween these two cities is 1920 km larger than 
listed in the standard references. Alternatively, the earth is round.

\subsection{Cartographic absurdities}

There are thus three glaring discrepancies between the finite disk/Gleason map and measurements on 
the globe:
\begin{enumerate}
\item The southern disk hemisphere has three times the area of the northern disk hemisphere 
[And no one noticed?]

\item The circumference of latitude circles monotonically \emph{increases} south of the equator on 
the disk, but monotonically \emph{decreases} on the sphere. [And no one noticed?]

\item The pole-to-equator distance is only 64 \% of its spherical value in the small disk model;
alternatively, the circumference of the equator is 57 \%  larger-than-spherical in the large disk model:
it is impossible to match both $R$ and $E$ on the disk to the generally accepted values on the globe. 
 [And no one noticed?]
\end{enumerate}

%We make the following assumptions
%\begin{enumerate}
%\item 
%\item
%\end{enumerate}

\clearpage   

\section{The Effective Gravity, Centrifugal Force and the Oblate 
Spheroid Approximation}

The rotation of the earth about its axis 
generates a centrifugal force that acts on the solid earth, oceans and all objects on the planetary surface. The centrifugal force acts  \emph{radially outward} from the rotation axis.

Gravity also acts on all that is subject to centrifugal force. Gravity is a vector pointing at the 
center of mass of the planet. 

\begin{figure}[h]
\centerline{\includegraphics[scale=0.8]{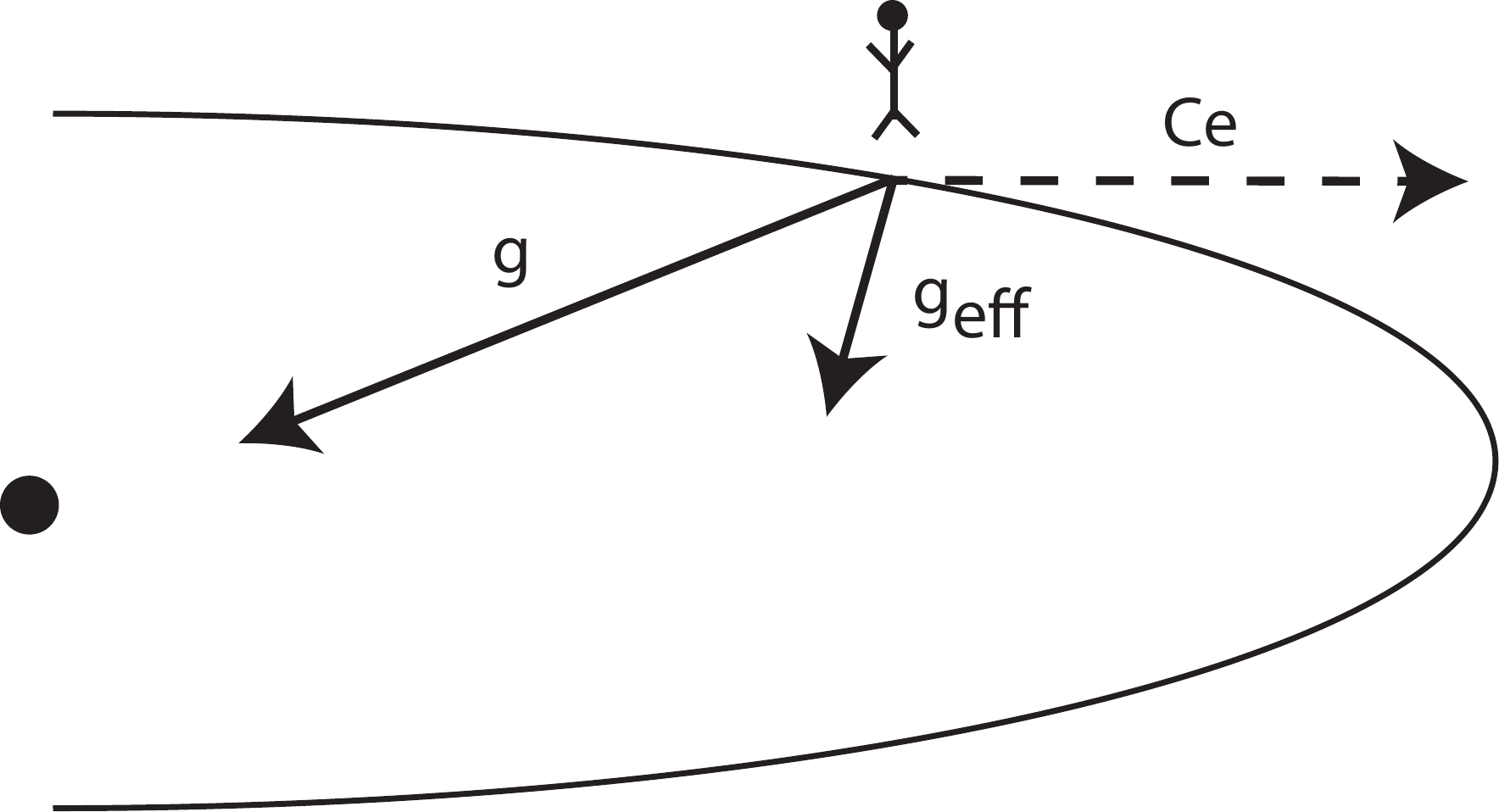}}
\caption{Cross section of half of an  oblate
   spheroid. The eccentricity is greatly 
exaggerated for graphical clarity. The stick figure
is an observer at a point on the planetary surface. At that point (and everywhere on the surface of the spheroid), gravity is a vector pointing at the 
center of mass, here shown as the black disk. 
The dashed line is the centrifugal force,which is everywhere radially outward from the rotation axis. 
The effective gravity $g_{eff}$ is the vector sum of the gravity vector and the centrifugal force 
vector. This vector is everywhere perpendicular to the surface of the oblate spheroid.}
\label{FigGravityEffective}
% Drawn by me in Illustrator 
\end{figure}

The combined effect of gravity and centrifugal
 force is the vector sum of the gravity and 
centrifugal force vectors. Meteorologists 
denote this sum as the ``effective gravity" 
   $\mathbf{g}_{eff}$ as shown in 
Fig.~\ref{FigGravityEffective}.

For graphical clarity, the magnitude of the 
centrifugal force relative to gravity has 
been greatly exaggerated. In reality, the largest
 value of the ratio of centrifugal force divided 
by gravity, which occurs at the equator, is 
only 1/300. The diagrams of people flung off the 
earth by centrifugal force found in many flat 
earth videos are hilarious fantasy except to 
the innumerate. One observable consequence 
of centrifugal force due to the earth's rotation 
is that at the equator, one weighs (299/300) of one's weight at the North or South Pole.

The other observable consequence is that a sphere of homogeneous 
fluid is deformed into an oblate spheriod by rotation. The change is such that the \emph{effective 
gravity vector} is \emph{perpendicular} to the \emph{surface} 
of the spinning spheroid at every point on its 
   surface.
 
Newton himself predicted the oblateness as 
confirmed by an intensive surveying expedition 
led by Maupertuis in the 1740's. The mathematics 
of self-gravitating spinning liquid masses in 
  equilibrium was developed further by Maclaurin in 
the eighteenth century, Jacobi and Poincar\'{e} 
in the nineteenth century and ``Black Hole" Chandrasekhar, N. Lebovitz and R. A. Lyttleton in the twentieth century \cite{Chandrasekhar,Lyttleton}. Jacobi predicted that at a high rotation rate,
  the branch of oblate spheroids would bifurcate, and 
the stable shape is then a triaxial ellipsoid. The dwarf planet Haumea is a Jacobi ellipsoid with the three axes being roughly 2300 km, 1700 km and 1000 km.
 
 The Earth, of course, is not liquid (now!), but 
sufficiently large solid masses and solid/liquid 
masses will plastically deform in response to the 
effective gravity as if liquid.
 The flat earth diskworld is \emph{not} an
equilibrium of a self-gravitating mass!

 % **********************  FIGURE  ***********
% ***************************************

\section{Hydrostatic Equilibrium}

\subsection{Containment of the Atmosphere}

The atmosphere is both static and dynamic. The simplest approximation to  the static part, usually called the ``mean flow", is obtained by applying a running time average to each of the meteorological fields:
\begin{eqnarray} U \equiv \frac{1}{P} \int_{t-P/2}^{t+P/2} \, u dt
\end{eqnarray}
where $u$ is any  of the wind components or temperature or density or any other meteorological variable and where $U$ is the corresponding mean.  If the period $P$ is anything from a week to a month, the mean is approximately independent of $P$ and is the static part of the flow.To a first approximation, the dynamic part is the  difference between a field and its mean.

 To a very high degree of approximation, the mean is in hydrostatic equilibrium. This means that the force of gravity on a small volume or ``parcel" of fluid is balanced by an upward pressure gradient The vertical momentum equation simplifies to
\begin{eqnarray}
\dfrac{1}{\rho} \dfrac{\partial \, p}{\partial z} = - g
\end{eqnarray}
where $\rho$ is the density, $p$ is the pressure and $g$ is the gravitational constant.

A popular Flat Earth mantra is, ``Pressure cannot exist without a container; otherwise, the 
atmosphere would just whoosh away into outer space."

This is true but a physical barrier is unnecessary. The pressure force is a \emph{force}, so
it can only be contained by \emph{another force}, in this case \emph{gravity}.

Hydrostatic balance, combined with the ideal gas flaw and the observation that the temperature 
of the mean flow does not vary much with height, implies that the mean pressure must decrease 
  \emph{exponentially}   with increasing height above the ground. The air pressure and density are \emph{halved} 
for each increase in height of about 6 km. Thus, vacationers who go from sea level to a hike in the
Rocky Mountains at an elevation of 5 km often suffer altitude sickness until their bodies adjust. 
Most climbers of Everest have to carry bottles of oxygen because the pressure at the summit is only about 
a third of what it is at the surface.

We can imagine confining a tall column of air in hydrostatic equilibrium inside a box. What happens if we open 
the lid of the box? Nothing. Gravity and pressure gradient are balanced. There is no acceleration of air 
either up or down. What happens if we remove the sides of the container? Nothing. Equilibria are 
very boring.

The components of the dynamic part of the flow, which meteorologists call the ``wave", ``eddy" or ``perturbation" flow, are in hydrostatic equilibrium also. However, motions on the scale of an individual thunderstorm or smaller have  deviations from hydrostatic balance that, while small, are dynamically significant.

At large heights, the unbounded atmosphere does suffer interference from other celestial bodies, the solar 
wind and so on. Mercury once had an atmosphere, but long ago lost it to space. 
Similarly, Mars lost most of its atmosphere although a low density of carbon dioxide remains. However, Earth, Venus and the other planets of the solar system have stronger gravity and will retain (most) of their atmospheres for billions of years to come.
 Through a combination of Jeans escape, polar wind escape, charge exchange escape, photochemical escape, sputtering escape and hydrodynamic escape, the earth is losing about 3 kilograms of hydrogen, 50 grams of helium and small amounts of heavier elements per second. All should be well for at 
least a billion years when a combination of hotter sun, runaway greenhouse effect and 
atmospheric escape could steal away the seas, but that is a tale for another time.

The atmosphere does not stop at a fixed height; instead, it merely fades away. 

\subsection{The Myth of the Firehose Ocean}

The pressure in the ocean increases by one atmosphere (1 bar $\leftrightarrow$ 14.7 pounds per square inch (psi) $\leftrightarrow$ 101 kPa) for every ten meters of depth. The deepest part of the ocean 
is 10,984 meters in the Marianas Trench where the pressure is 1,071 atmospheres or 15,570 pounds per 
square inch. With so great a pressure difference between the water in the deep ocean and the 
pressure of (on average) one atmosphere at the sea surface, why doesn't the 
briny ocean erupt upward like the jet of an immensely powerful firehouse aimed at the heavens?

The answer for the sea is the same as for the sky: \emph{hydrostatic equilibrium}. A column of ocean water or air
an inch on each horizontal side and extending from the bottom of the Marianas Trench upward weighs 
15,555 pounds if the columns extends only to the ocean surface, or 15,570 pounds if the column extends 
100 kilometers into the sky. The weight of the column (15,555 pounds of salt water, 14.7 pounds of air)
is exactly equal to the pressure at the bottom of the column.

If the bottom of the column is moved upward, the weight becomes less and the pressure 
becomes less in exact proportion. The change of the pressure with depth is a vertical pressure 
gradient. The upward pressure force exactly balances the downward force of gravity to 
produce an equilibrium, devoid of movement up or down.

The equilibrium is not perfect; it is disturbed --- slightly -- by the surface gravity waves that rock vessels on the sea and other dynamical phenomena. Metaphorically, dynamics is a dance and 
hydrostatic equilibrium is the dance floor. The floor vibrates as the dancers tap and stamp, 
but the floor remains the foundation for all the motion above it.

\section{Solid Body Rotation}

\begin{quotation}
No, the
atmosphere is absolutely not velcroed to
 to the frame of reference with the
earth!
\end{quotation}
\hspace*{0.5in} --- Bob Knodel in minute 37 of \cite{GLobebustersDebunkingProfDave}.

The mean east-west velocity is a complicated function of latitude and height because of the transfer of 
momentum both latitudinally and vertically by waves. ``Wave-mean flow"  interaction is a topic 
too complicated to discuss here but see \cite{Buhler14,BoydOP1,Andrews85}. Furthermore, the mean flow   also contains the Hadley circulation. This is independent of time and therefore part of the mean, but it is a convective circulation, responding to spatially-varying heat sources, which is dynamic. Without wave-mean 
flow interaction and the Hadley cell, the mean is a state of ``solid body rotation". It is useful to look at this because, metaphorically, 
solid body rotation is the wet plaster wall on which the wave-mean interactions and Hadley flows are painted and absorbed, or if one prefers a different metaphor,   the dance floor on which the dynamics dances. 
 Furthermore, an improved approximation of  the static part of the atmosphere subtracts from the mean  the wave-mean flow 
effects, which are dynamic.

Solid-body rotation is a static, hydrostatically-balanced flow such that the velocity of 
the rotating atmosphere at a given point is identical with that of a solid body at the same point rotating
 with the same angular velocity.
 The velocity is a function only of radial distance to the center of the sphere.

 Viscosity transmits shearing stress from one layer of fluid to another, in this case,
  from one radius to another. Solid-body rotation is equivalent to zero stress and is therefore 
the end of this process of viscous adjustment.

It is easy to demonstrate this in a student lab. Fill a cylindrical dish with liquid and place it on a turntable. 
When the electric motor is turned on, the fluid in a 
thin layer (``boundary layer") accelerates very rapidly  to match the velocity on the rim of the dish; 
the fluid over most of the dish is motionless. As time passes, the boundary layer thickens and thickens,
evolving to a state of solid-body rotation in which the tangential velocity is a linear function of the radial distance from any point in the flow to the center of the dish.

Knodel's assertion is wrong; the solid,  spinning sphere controls the static atmosphere. Its ``velcro" 
is a combination of gravity (for hydrostatic balance) and viscosity (for transmission of shearing stresses and for evolution to solid-body 
rotation).

\section{Hadley Cell}

The equator is hot; the poles are cold. This drives a convective circulation
 with hot air rising at the equator and cold air sinking elsewhere to complete the flow.
 The circulation is independent of longitude (``axisymmetric") if the continent/ocean asymmetries are ignored, which is a useful 
first approximation. The name comes from 
George Hadley (1685-1768), an English lawyer who published a four page article
 describing and explaining the flow in 1735. Hadley recognized that on a rotating earth,
  the Coriolis force would apply a westward deflection to the surface  flow converging  towards the equator 
in both hemispheres. This westward (``easterly") flow is the ``trade winds".

On the flat disk or on a slowly-rotating spherical planet, the Hadley cell is global with the descending 
flow at the poles. However, our planet spins so rapidly that the Hadley flow does not reach the poles. Instead,
 the air descends in the subtropics in bands centered at $\pm 30^{0}$ latitude.

The ocean temperature in equatorial regions is very warm, so the 
evaporation rate is high,  forcing 
 the surface air to be nearly saturated. As this equatorial air rises in the Hadley updraft, the rapid decrease 
of pressure forces the air parcels to expand. The ascent is ``adiabatic" in the sense that parcels exchange 
little energy with their surroundings. The adiabatic expansion generates ``adiabatic cooling". The lowered 
temperature reduces the capacity of the air to hold water as vapor. Saturation occurs, meaning that 
the relative humidity is 100 \% and condensation into liquid droplets follows. The phase change generates latent heat 
release which warms the air and makes it buoyant, further driving the moist air upward. The liquid water 
forms clouds and rain. The heavy, steady rain makes the equatorial region the  
\emph{tropical rain forest}.

When the air descends in the subtropics, it is rather dry since much of the water vapor has condensed in 
the upward flow at the equator, and rained onto the equatorial jungles. The descending air parcels are compressed and experience 
``adiabatic warming". This warming increases the carrying capacity of the descending air, so the 
relative humidity decreases sharply even though the absolute amount of water vapor in the 
air does not change.

The world's major \emph{deserts} --- the Mojave in the U. S., the Gobi in Asia, the Sahara and Kalahari in Africa, etc. ---
lie in the \emph{descending} regions of the Hadley flow. 

Of course, the subtropics are not uniformly desert, and there are also local deserts  outside the subtropics. Mountain ranges and local wind and ocean circulations modify
the global dynamics. Even so, condtions for deserts are very favorable in the dry, descending Hadley flow.

The Hadley circulation is extremely efficient at redistributing heat. The mean atmospheric temperature 
is almost uniform over the whole domain of the Hadley cell. There is little variation in mean surface 
temperature over the entire tropical band on earth. Venus is very slowly rotating, so the Hadley circulation extends to the poles; the mean surface temperature on Venus has, compared to earth,  a very small latitudinal gradient.

On a flat earth, the Hadley flow should be Venus-like. The Arctic and Antarctic regions would be only be 
slightly cooler than the tropics. Florida and Calgary would differ little in temperature. 

Also, on a flat, non-rotating earth, the trade winds disappear! Goodbye, Columbus!

 % **********************  FIGURE  ***********
\begin{figure}[h]
\centerline{\includegraphics[scale=1.0]{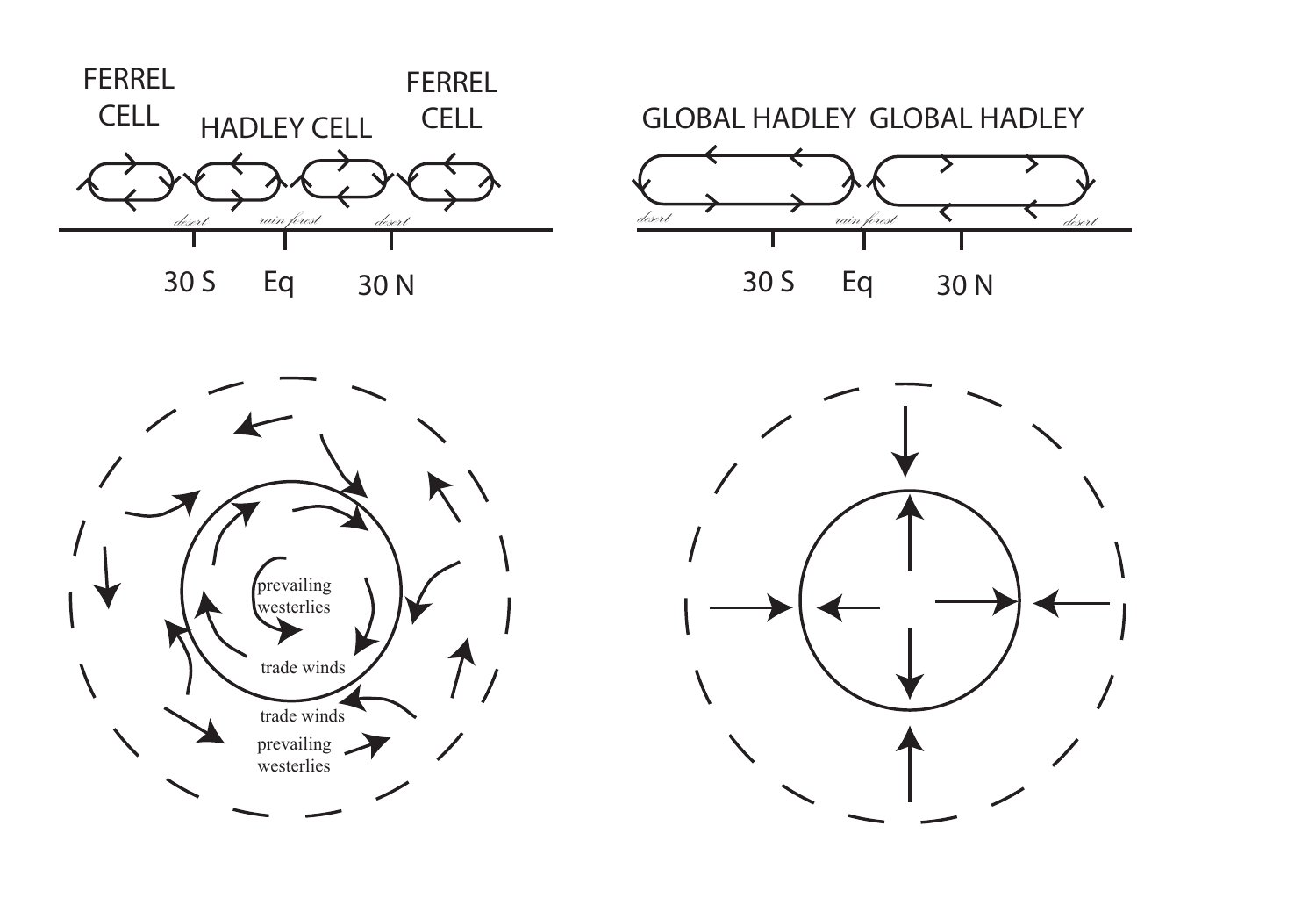}}
\caption{Schematic of mean, longitude-independent surface  flow on the globe (left)  and flat earth (right). Note that the solid circle is the equator and the dashed circle is the South Pole. 
 The top graphs are height-latitude cross-sections. The bottom plots are views looking down from above the North Pole.}
\label{FigFlatFlow_GlobeFlow}
% Drawn by me in Illustrator 
\end{figure}  
% ***************************************

\section{Midlatitude Synoptic Vortices and Baroclinic Instability}

The latitudinal  bands on our planet, poleward  of the tropics are, on our fast-spinning planet,
occupied not by the Hadley cell, but a circulation cell of the opposite rotation called the ``Ferrell cell". 
Because the axisymmetric flow is equatorward at the surface in the Ferrell cell, Coriolis force
deflects the winds eastward, generating the ``prevailing westerlies". This part of the axisymmetric flow
is strongly varying in both latitude and height. It is unstable to small amplitude waves which grow and 
grow at the expense of the Ferrel cell. This spontaneous growth of small waves into big vortices is called ``baroclinic instability". It was first
explained  by Jule Charney in 1947. (He employed a tangent plane approximation, the so-called ``beta-plane".)

The waves eventually roll up into vortices. Low pressure centers alternate around a latitude circle 
with high pressure centers. The low pressure centers are ``cyclones" in either hemisphere, but spin,
as viewed from above, counterclockwise in the Northern Hemisphere but clockwise in the Southern Hemisphere.
 The rotation is in the opposite sense for high pressure centers, which are ``anticyclones". 
 Collectively these are ``synoptic-scale vortices". 

These vortices slowly grow and decay, but to within an error of about 10 \%, the pressure force 
directed inwards (for cyclones) or outwards (for high-pressure centers) is balanced by Coriolis force.

Hurricanes and typhoons are vortices that always begin in the tropics and are always 
cyclones. Again, the sense of rotation in the southern hemisphere is opposite that of the 
northern hemisphere. The dominant dynamics is again quasi-steady state with a 
balance of Coriolis and pressure forces. In the eyewall region where the tangential 
winds are strongest, however, centrifugal force is as important as Coriolis force.

On a flat earth without spin, baroclinic instability would disappear, and with it, most of what 
we in the middle latitudes think of as the weather. Fronts, squall lines, and tornado-spawning 
  mesoscale systems would be missing, too.

\subsection{Dishpan Laboratory Experiments}

\begin{quotation}
``Approximation is as important as derivation"
\end{quotation}
\hspace*{0.5in} Thomas B. W. Kirk (Harvard lecture,1970)

\bigskip

Dave Fultz showed that baroclinic instability could be captured in the laboratory by taking a cylinder or annulus
of water in a dish on a rotating turntable. (A spinning cylindrical container is a ``dishpan" in
geophysical jargon.) The flow represents a single hemisphere; a temperature 
gradient in latitude (radius in cylindrical coordinates) is created by heating the outer rim (``equator") and 
cooling the center of the cylinder (``arctic"). The flow is visualized by sprinkling aluminum powder on the 
surface of the water. When the apparatus is started from rest and the cylinder spun up to a constant rotation rate, a Hadley-like flow 
independent of the polar angle $\theta$ in polar coordinates emerges first. When the flow is  sufficiently strong,
 baroclinic instability sets in, and wavy perturbations  spontaneously amplify, stealing energy from the
axisymmetric ($\theta$-independent) flow. The waves become vortices qualitatively resembling 
 synoptic-scale cyclones and anticyclones in the atmosphere \cite{SpenceFultz77,RiehlFultz58}.
Beautiful illustrations are in Van Dyke \cite{vanDyke82}. 

Allaby \cite{Allaby95} illustrates a recreation of  Fultz' experiments by middle school students.
 The turntable is a cake dish, spun by hand. The heat source is a set of candles underneath the 
outer wall of the dish. The cold source is a cylindrical can filled with ice cubes and placed in the center of 
the dish. Paprika is sprinkled on the surface of the water to visualize the flow.

\subsection{Microgravity Laboratory Experiments with Spherical Symmetry}

The cylinder can be replaced by a sphere by (i) using a dielectric fluid and mimicing gravity by a 
spherically-symmetric  
electric force created by charging the sphere and (ii) eliminating the non-spherically-symmetric force of  gravity by 
performing the experiment in free fall (``microgravity").
 The 1985 experiments on the shuttle \emph{Challenger} used an apparatus built by John E. 
Hart at the University of Colorado \cite{Hartetal86,HartGlatzmaierToomre86}.
 These experiments have continued on the International Space Station (ISS) \cite{Futteretal13,Zaussingeretal19}. 
Experiments on fluids in microgravity have also been performed using aircraft flying  a 
parabolic trajectory like NASA's famous ``Vomit Comet". Such flights only provide  30 seconds  of freefall per dive, but this is long enough for some flow experiments \cite{Meyeretal19}.

\subsection{Soap Bubble Vortices}

\begin{figure}[h]
\centerline{\includegraphics[scale=1.0]{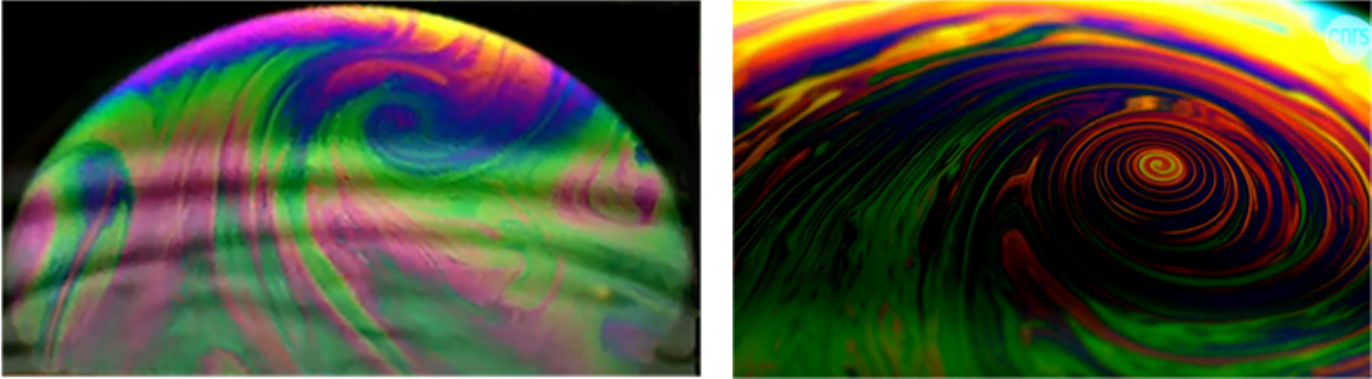}}
\caption{Left: a thin film uniformly covering a sphere (soap bubble). Right: close-up of a soap bubble
vortex.}
\label{FigVortex_on_Half_Bubble_2plot}
 % Screenshots, two assembled into one figure, from \cite{SoapBubbleCyclone}
\end{figure}

A group at the  ``Ondes et Matière"  CNRS laboratory in Aquitaine (France) performed  laboratory experiments on vortices in a soap bubble with spherical geometry \cite{Meueletal13}. 
  The solid earth is replaced by  a sphere of air; the ocean or atmosphere is modeled by the thin liquid film wrapped uniformly 
around it as illustrated in Fig.~\ref{FigVortex_on_Half_Bubble_2plot}.

 % **********************  FIGURE  ***********
% **************************************

Thus, the claim that meteorology and physical oceanography are strictly \emph{observational sciences}  
 is not entirely true. It is possible to do instructive laboratory experiments even in spherical geometry.

To be sure, it is not possible for laboratory experiments to include all the radiative transfer, photochemistry,
ice physics, etc., included in general circulation numerical models. Thus, modern meteorology is even more so a
  \emph{computational} science.

\section{The Beta Effect}

\subsection{Rossby Waves}

One deficiency of Fultz' spinning cylinders is that the Coriolis parameter is a \emph{constant}, equal to twice
the angular frequency $\Omega$ of the rotating apparatus. On a sphere,
the Coriolis parameter $f$ is $2 \Omega \sin(\text{latitude})$. It thus is a maximum at the North Pole,
diminishes to zero at the equator and is negative everywhere in the southern hemisphere.

The constancy of the Coriolis parameter in experiments with a 
flat-bottomed cylinder is unfortunate because
 many geophysical phenomenon depend on the 
rate  at which the Coriolis parameter is varying. By convention, the latitudinal derivative of the Coriolis parameter is always denoted by the Greek letter ``beta" [$\beta$]. In the early 20th century, planetary scale waves retrogressing westward relative to the meanl flow were discovered observationally. Carl-Gustav Rossby and Jacob Bjerknes independently provided a theoretical explanation in 1937. These waves, always with the phase velocity westward relative to the mean flow, are now called 
``Rossby waves" \cite{Platzman68}. Observations  confirm theory in both the ocean and the atmosphere.

Madden \cite{Madden07} is a good review of atmospheric Rossby normal modes. 
The most prominent of these is the so-called ``five-day wave", which propagates westward at 90  m/s, completing one full rotation  of the globe in five days. This mode has longitudinal structure proportional to 
$\sin(\lambda)$ where $\lambda$  is longitude in radians. High pressure at the Greenwich meridian is mirrored by low pressure at the international dateline. 

Similarly, some modes of atmospheric tides are Rossby waves \cite{ChapmanLindzen70,BoydOP239,BoydOP254}.

High-tech equipment is not required; a barometer and a personal computer (to compute 
Fourier transforms) are sufficient to detect these kind of periodicities in atmospheric flow.

\subsection{Stabilizing Long Waves in Baroclinic Instability}

The beta effect is not essential to baroclinic instability, or else Fultz' pioneering experiments with a 
rotating cylinder would have failed. However, the beta-effect does significantly modify baroclinic instability by \emph{stabilizing} the long waves, that is, the waves of largest spatial scales. The result is that on the globe, only waves of intermediate scale are unstable. The horizontal scale of these unstable waves is roughly equal to the so-called ``Rossby Deformation Radius" which depends on the strength of the Coriolis parameter. 
Without the stabilizing effect of nonzero $\beta$, midlatitude storms would be much larger in scale, and therefore more damaging.

\subsection{The Gulf Stream}

The Gulf Stream is a strong narrow current that runs along the east coast of the United States 
from Florida to Cape Hatteras, North Carolina, where the jet turns away from the coast, develops
wavy instabilities and finally evolves to intense, circular vortices called ``Gulf Stream rings".  
 
The Gulf Stream has been known to sailing masters for many centuries.  The first scientific investigation was by Ben Franklin. On his trip back to America in 1775 after his long exile in England, his grandson repeatedly dropped a bucket into the ocean, retrieved it and measured the water temperature with a mercury thermometer; the Gulf Stream is much warmer than the cold North Atlantic and thus can be identified and mapped by temperature alone. Franklin did not publish his map of the Gulf Stream, coauthored with his cousin, a merchant captain, until after the American Revolution. (He was rather  busy during the war as the rebels' ambassador to  France, successfully negotiating French intervention.)

The Gulf Stream is but one instance of a ``western boundary current". Others include 
the Brazil Current (South Atlantic), the Somali Current (North Indian Ocean), the Aghulas Current
(eastern coast of Africa), and the Kuroshio Current (North Pacific). Why? And why are there no 
\emph{eastern} boundary currents?

After two centuries of failure and futility, the true explanation was given by Henry Stommel in  1948. Science has its urban legends; an oceanographic tall tale is that Stommel was so pleased with his Gulf Stream discovery that he danced a jig  in the sedate hallways of the  Wood Hole Oceanographic Institution. 
Many years later, Stommel confirmed to my colleague (and his old friend) Stan Jacobs that the story was in fact true. He was later awarded the Crafoord Prize [the geophysics Nobel Prize]\footnote{The Crafoord Prizes were funded by Dr. Crafoord to provide Nobel prizes in areas of science which were not specified in Alfred Nobel's will. The Crafoord  Prizes are administered by the same organization that awards the Nobel prizes. This body previously instituted the economics Nobel prize, funded by the Bank of Sweden, so Crafoord expected no difficulties. However, the Academy took so much criticism for extending the prizes into the social sciences that instead of renaming the economics prize after the Bank of Sweden and instituting new Nobel prizes in science, the Academy continued the ``Nobel" Prize in economics while administering the new science prizes under the funder's name, despite his opposition.},  though not for dancing.\footnote{Among his many achievements, Stommel coauthored a book entirely devoted to the 
Coriolis force \protect\cite{StommelMoore89}.}. 

Stommel showed that the beta-effect not only makes Rossby waves propagate westward, it also squeezes jets against the western boundary of the ocean \cite{Stommel48}. The Gulf Stream and all its western boundary current brethren would not exist on a disk, not even on a rotating flat plane.

\subsection{Sverdrup Flow}

In a famous paper, Harald Sverdrup (1947) \cite{Sverdrup47} developed a simple formula, now named for 
him, for the depth-averaged north-south mean velocity as driven by wind stress.

Let $\beta$ denote the latitudinal derivative of the
Coriolis parameter. Let $\rho$ be density, assumed constant [an excellent oceanic approximation]. Further 
identify 
$\overline{v}$ as the mean north-south depth-averaged
oceanic current. Let $\tau^{x}$ and $\tau^{y}$ be the
east-west and north-south components of the 
surface wind stress. Lastly, denote the unit vector in the vertical by $\hat{k}$.
If we integrate over depth, assume that the stress is zero at the ocean bottom, 
replace the divergence of the horizontal velocity by $\partial w/\partial z$ 
and assume the vertical velocities are negligible at the boundaries\footnote{The vertical 
velocity is not negligible because of surface gravity waves, but 
 after we take a running time average or space average to filter the small-scale 
surface waves, the \emph{averaged} vertical velocity is negligible.}, 
we obtain
\begin{equation}
\beta \overline{v} = \frac{1}{\rho} \left( \frac{d \tau^{(y)}} {dx}
 \, - \, \frac{d \tau^{(x)}} {dy} 
\right) = \frac{1}{\rho}  \left( \hat{k} \times \vec{\mbox{curl}} \vec{\tau} \right)
\qquad \mbox{[Sverdrup Relation]}
\end{equation}

The trade winds and prevailing westerlies create a wind stress with non-zero curl, which results in 
a clockwise gyre in the north Atlantic. In the 
absence of the beta effect, the Gulf Stream would be 
a broad, slow-moving current instead of the narrow jet
  that is actually observed. The return, southward flow \emph{is} broad and slow-moving; its dynamics are described by the Sverdrup relation.

A flow everywhere equatorward would strip the ocean 
dry at high latitude and pile up mass at low latitude. Thus there 
must be a region where the Sverdrup relation is inaccurate.
  In a narrow boundary 
layer along the western side of the Atlantic, a different balance of forces prevails. In this boundary layer, 
  the Gulf Stream is the northward return flow 
of the gyre.

Note that the Sverdrup flow is inversely proportional
to $\beta$. Without a Coriolis force, and further,
one in which the Coriolis  parameter varies with 
latitude, $\beta=0$ and the Sverdrup flow does not exist.

\section{Ekman Boundary Layers}

On the voyage of the Arctic exploration steamship \emph{Fram} from 1893-1896, the Norwegian scientist and explorer F. Nansen noticed something very odd. Near the Arctic icepack, floating bits of ice provided a natural flow visualization. The wind exerts a stress on the ocean and drives a surface current. The oddity 
was that the surface current vector made an angle 
twenty to thirty degrees to the right of the wind vector.

Nansen's doctorate was in marine neuroscience and he lacked the mathematics to make a theory. He was, however, a very good scientific evangelist, and he described this non-parallelism of wind and 
surface current to any geophysicist who would listen. V. Ekman published a simple theory in 1904.

The Ekman flow is a boundary layer current whose 
magnitude falls exponentially with depth. There is a three-way balance of forces between the wind 
stress, Coriolis force and viscosity in the Ekman layer. It is the Coriolis force that causes the nonzero angle between the wind stres and the surface current. The angle is 
forty-five degrees in Ekman's idealized theory;
 more accurate theories give a somewhat smaller angle as observed by Nansen.

A remarkable feature of oceanic  Ekman layers is that the current vector rotates clockwise in the northern hemisphere with increasing depth. A few dozen meters down, the Ekman current is directly \emph{opposite} 
in direction to  the wind stress that drives it!

Ekman's solution, with different boundary conditions, applies also to the atmosphere. 
In the atmosphere, the Ekman flow is driven by 
the geostrophic flow\footnote{``Geostrophic flow" is shorthand for ``geostrophically-balanced flow", that is a flow in which the dominant terms in each of the two 
horizontal momentum equations are pressure gradient force and Coriolis force.} at higher altitudes. The Ekman flow a kilometer up is parallel to the geostrophic 
flow. The wind vector rotates counterclockwise with decreasing height so that the surface wind is to the left of the direction of the geostrophic flow above the boundary layer by an angle that is 45 degrees in Ekman's theory and 20 to 30 degrees in observations and improved theories.

 All rotation 
directions must be reversed in the southern hemisphere. Ekman theory is singular at the equator where additional terms in the Navier-Stokes equation must be retained to accurately model the observed equatorial boundary layer.

Ekman theory does not replace the usual theories 
for hurricanes and synoptic-scale cyclones and anticyclone. Rather,
Ekman theory is a description of the boundary layers that form in the lowest kilometer of the 
atmosphere. Above the boundary layer and up to the tropopause, hurricanes 
and synoptic cyclones are geostrophically balanced
 as noted earlier.

The Ekman layer underneath a hurricane or synoptic-cyclone is an ageostrophic flow that sends
air spiraling in towards the center of the vortex 
where it turns the corner and erupts out of the 
boundary layer as a strong vertical flow. (This is called ``Ekman  pumping").  For a
hurricane, the boundary layer radial inflow does not reach the center because of 
centrifugal force, which grows as the spiraling blobs 
of fluid get closer and closer to the center. Instead, the boundary layer inflow erupts in an annulus which is called the ``eyewall". The upward flow triggers violent
 convection which provides the energy to sustain the 
entire vortex. Over the tropical ocean where hurricanes grow, the boundary layer fluid fed to the eyewall is very moist and warm, perfect fuel for a monster storm.

Boundary layers without Coriolis force are common; the
flow over an airplane  is a familiar example. However, for
  Ekman boundary layers and Ekman pumping, Coriolis force is as essential as gills are essential to a fish.

% \section{Geostrophically-Balanced Currents, the % Thermocline and Two-Layer Models}

% TO BE WRITTEN

\section{Bubbles}

Blowing soap bubbles is a familiar childhood activity. Who knew those semi-transparent spheres, 
floating in the air, are models of our  planet?

\begin{figure}[h]
\centerline{\includegraphics[scale=0.5]{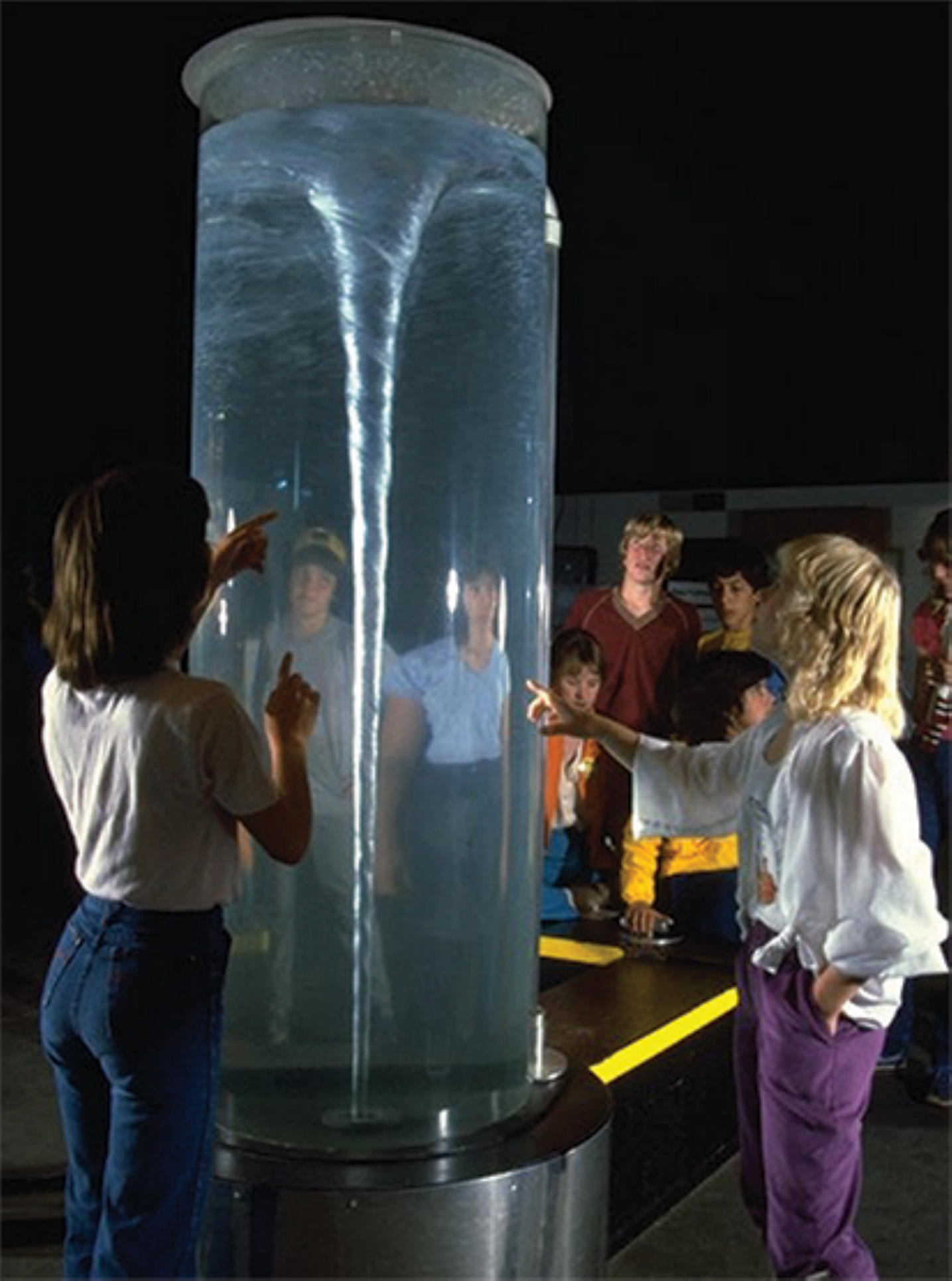}}
\caption{A steady draining vortex, also called a ``free vortex" or ``line vortex" at the museum known as the San Francisco Exploratorium.}
\label{FigExploratoriumFreeVortex}
 % cropped version of image copyright 1194 by the Exploratorium
\end{figure}

A favorite trope of   Flat Earth believers is ``Water always seeks 
its own level". This is interpreted as a [spurious] physics law that the [mean] ocean surface 
is a plane (and therefore the planet itself is a flat-sided disk). 
This steady-flow-is-always flat ``law"  is falsified by a variety of 
classical steady flows such as  the draining vortex (Fig.~\ref{FigExploratoriumFreeVortex}), 
hydraulic jumps in rivers and soap bubbles (\cite{Meueletal13} and  Fig.~\ref{FigVortex_on_Half_Bubble_2plot}). 
   In addition, the mean surface of the ocean is not flat as explained in the next section.

A soap bubble is a sphere of air surrounded by a thin film of liquid, held to the surface by 
surface tension. The earth is a sphere of rock and itron 
surrounded by a thin film of salt water, held in  place by gravity.

The notion that the ocean surface must always be flat is a Failure of Imagination. 
 Sometimes the best refutation of a wrong idea is mathematics. Sometimes, 
a better refutation is to blow bubbles.

\section{Ocean Surface Topography and Thermocline/Pyncocline Topography}

The ``geoid" is that particular equipotential surface of the 
\emph{effective} gravity which conforms most closely to the mean surface of the ocean, also known as  ``mean sea level". The mean  is what remains after surface waves  have been scrubbed 
away by a running time average. The ``effective gravity" is the sum of gravity and centrifugal force as defined earlier.

 % **********************  FIGURE  ***********
% **************************************

\begin{figure}[h]
\centerline{\includegraphics[scale=0.5]{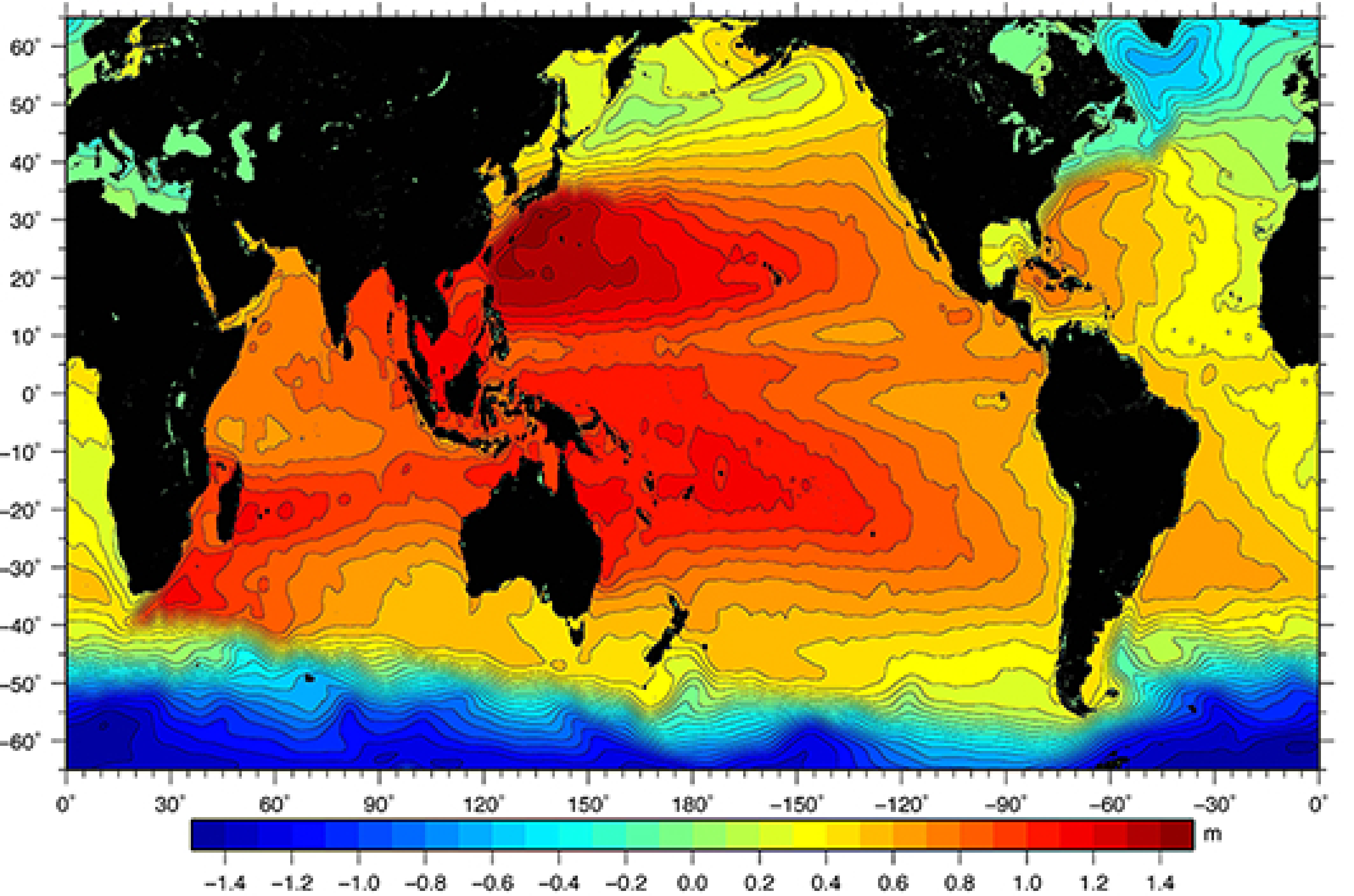}}
\caption{Ocean surface topography.}
\label{FigOceanTopo}
% NOAA r 
\end{figure}

The TOPEX/Poseidon, Jason-1 and Jason-2 satellites, launched in 
1992, 2001 and 2008,   have mapped the topography of the ocean 
surface through radar altimeters. The deviations 
of the mean from the geoid are only a maximum of a meter and a half,
but Jason-2's altimeter error is only about 3 cm.
    This is sufficient to capture an impressive collection of hills, valleys 
and slopes as illustrated in Fig.~\ref{FigOceanTopo}.

 % **********************  FIGURE  ***********
% ***************************************

The ocean, like the atmosphere, is hydrostatic equilibrium. The pressure 
in a water column at a given depth is proportional to the weight of all the fluid above that depth in the column. A hill in mean ocean surface 
topography is a center of high pressure while a valley is a pressure low.

What balances these pressure gradients? For small, intense 
ocean vortices, the dominant balance is centrifugal force and horizontal pressure gradient force, a ``cyclostrophic" balance. 
 For features with a horizontal length scale of a hundred kilometers or 
larger, scale analysis shows that Coriolis force balances pressure force.
 Thus, the satellite altimeter maps, Fig.~\ref{FigOceanTopo}, show us that geostrophically-balanced jets and vortices are as ubiquitious in the ocean as 
in the atmosphere.

\clearpage 

\section{The Rossby Radius of Deformation and the Geostrophic Adjustment Problem} 

A favorite game of children of a certain age is to make an expanding 
disk of water waves by tossing a rock into a lake. A  few moments 
after the impact, radial expansion and viscosity diminish the 
waves to nothing, and the water surface is flat. (Bristol Lake in
 New Hampshire  and Clare Lake in Michigan, my favorite childhood targets, have little surface 
wave activity on calm days except when a power boat zooms by.)

Carl-Gustav Rossby recognized long ago that meteorological instruments can fail or otherwise give wildly erroneous 
measurements. The effect of a bad 
observation is  a point perturbation of 
the weather forecast or of the ocean surface, mathematically equivalent to throwing a rock 
into a lake. Buoyancy waves will radiate away from the point, as on a 
lake, but on a rotating planet, the flow does \emph{not} relax 
back to a flat surface, but rather to a \emph{geostrophically-balanced}
time-independent flow. This is final state of the ``Rossby geostrophic adjustment 
problem". Many of the finest minds in geophysical fluid dynamics have 
refined Rossby's idealized, approximate analysis. 

Many of the stationary and traveling bumps in the ocean surface topography are the steady currents and traveling waves excited 
by adjustment of the ocean to localized wind stresses, such as those 
beneath a hurricane.

The characteristic horizontal length scale for  ``geostrophic adjustment" is the ``Rossby radius of deformation" $R_{D}$. This turns out
to be a fundamental length scale for both the ocean and the 
atmosphere. For disturbances whose spatial scale is small  
 compared to the Rossby radius, Coriolis force is unimportant whereas 
for disturbances whose horizontal scale is large  compared to $R_{D}$, 
 Coriolis force is dominant. The synoptic-scale vortices that 
dominate weather in the middle latitudes as well as ``Rossby adjustment scars" have length scales roughly equal to the Rossby 
radius.

Parenthetically, note that the Rossby radius in the ocean is only 
about one-tenth its value in the atmosphere. This means that 
an ocean forecasting model must have one hundred times the number 
of numerical degrees of freedom (``grid points") as an atmospheric
weather prediction model of comparable accuracy.

By coincidence, the Rossby radius for Mars is about the same as 
for the terrestial atmosphere, but the Martian planetary circumference is 
much smaller than on earth. The dominant baroclinically unstable 
waves  on Mars have three crests and troughs around a circle 
of latitude versus roughly six maxima and minima around a 
latitude circle. Thus, forecasting Martian weather would be easier 
than on earth, if sufficient observational data were available.

The greatest difficulty for both ocean prediction and Martian weather 
forecasting is obtaining high resolution data to initialize the model.    
 In the latter half of the next section, we discuss a new observational tool that has 
vastly increased ocean data.

\section{Two-layer Models or Fun with Oil and Water}

Geophysical fluid dynamicists are like artillery officers: the first shots 
are hopefully close but usually not quite on target. With successive 
corrections (from forward observers or spotting aircraft or in fluids through 
higher order in perturbation series or adding additional small terms), 
the salvos are walked into the target.

The opening salvo in the approximation of the ocean is to assume a 
layer of incompressible fluid of constant temperature and salinity (``homogeneous fluid"). A far 
more realistic approximation is to allow the flow to be continuously 
stratified and fully three-dimensional. A useful intermediate is the ``two-layer model". This pretends that the ocean is composed of two immiscible 
layers of homogeneous fluid of different densities, such as oil and water.

 % **********************  FIGURE  ***********
\begin{figure}[h]
\centerline{\includegraphics[scale=0.3]{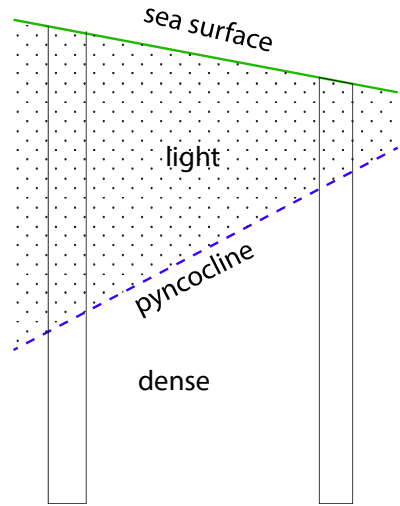}}
\caption{Schematic of two-layer model of the equatorial Pacific Ocean. The upper layer (textured) is
light and hot; the water below the pyncocline (dashed blue line) is heavy, cold abyssal water (unshaded). 
In the laboratory, the upper layer would be oil and the lower would be water.
 During normal non-El Ni\~{n}o conditions as shown, the mean sea surface is tilted so that it is highest 
in the west [left side of diagram]. The tilt is driven by the wind stress of the trade winds, which applies a westward force to the 
upper layer. The pyncocline tilts in the opposite direction from the surface but with typically 200 times the slope of the sea surface. 
 The thin black vertical lines bound two columns of water. The hydrostatic pressure at a point on the sea bottom 
 is the weight 
of the column of water above that point. If there are to be no horizontal pressure gradients in the abyss, then the
two  columns must have equal weight. The left column is taller than the right column because the sea surface is tilted.
 The opposite and much greater slope of the thermocline compensates so that the shorter, rightmost column has less water but a much 
higher percentage of dense, cold absyssal water. The leftmost column, despite its greater height, weighs no more than the other 
column because the leftmost column has a much greater volume of light, hot upper layer water. For graphical clarity, the drawing is
not at all to scale, and the ratio of surface slope to pyncocline slope, typically 1/200, has been greatly exaggerated.}
\label{FigTwo_Layer_Tilted_Thermocline}
% Drawn by me in Illustrator as Two_Layer_Tilted_Thermocline.ai
\end{figure}  
% ***************************************

The two-layer model is surprisingly useful  because to a first approximation, especially in the tropics, the ocean \emph{is} a 
two-layer system. The upper fifty or a hundred meters of the ocean 
is mixed into a uniform density by turbulent convection driven by 
solar heating. Under this ``mixed layer", the abyssal flow is cold 
and dark; it varies little in density. The boundary between abyssal 
flow and the surface mixed layer is a thin layer of  high gradient in
both density and temperature. The terms ``pyncocline" and ``thermocline" are applied as near-synonyms with the word choice 
depending on whether one is displaying a graph with density contours or temperature 
isolines. The two-layer model sharpens the thermocline to a jump 
discontinuity. In the ocean, the density difference $\delta \rho$ between the 
two layers is usually half a percent or less, but the dynamical 
consequences of the second layer are profound.

I was first exposed to the wondrous fairyland of two-layer models
when I visited Woods Hole Oceanographic Institution in 1973
  and Jack Whitehead gave me a tour of his laboratory where half a dozen experiments were going simultaneously. Waves can propagate 
along the oil-water interface. These are ``baroclinic waves" or ``waves in the first baroclinic vertical mode".  If the density difference between layers is small, as typical of the sea,  the speed of these intrafacial waves is very slow compared to waves on the oil-air interface\footnote{Most oils are less 
dense than water, so the oil layer floats above the water layer without 
mixing or dissolving into it.}. Indeed, as I watched jets plunge from a 
shallow shelf into the abyss and waves run around the edges of a flow basin, it seemed that everything was happening in slow motion, but
without the intermediary of a video camera.
 
  All of the experiments were on turntables, rotated by speed-variable electric motors. All bow before the Lord Coriolis, who rules the 
  oceans and the winds! At least phenomena on the Rossby deformation radius or larger scale.

Ron Smith (Yale) once enlivened and amused an American Meteorological 
Society conference by demonstrating the generation of mountain waves using materials purchased entirely at a hobby shop except for a gallon jug of yellowish corn oil bought at the local supermarket. The fluid was confined between the glass panels of an ant farm. The recirculating mean flow was generated by an acquarium pump. The ``mountains" were a block of wood, just thick enough to fit in the three centimeter gap between the two large panels of the ant farm. The upper boundary of the block was machined to have curving 
hills and depressions
  that forced the flow to oscillate up and down as it flowed past, 
generating stationary waves in the process. 

The two-layer model is closely connected with the ocean surface 
topography because, assuming there are no horizontal 
pressure gradients at the bottom of the sea, the shape of the 
thermocline must be a \emph{reverse image} of the ocean surface,
 \emph{amplified}  by the huge factor $1/\delta\rho$, the fractional density 
difference between the layers. The reason 
is that if an isobath (surface of constant pressure) is horizontal at 
some depth, then different columns of seawater must have the same 
total weight on that isobath. (See Fig.~\ref{FigTwo_Layer_Tilted_Thermocline}.) Assume $\delta\rho = 1/200$, a typical value. If the ocean surface is one meter above the geoid at one 
point but even with the geoid at a different location, then the column
under the hill must be lightened by replacing 200 meters of 
dense, lower layer water by lighter mixed layer fluid. This implies that, under the hill,
the thermocline must descend  two hundred meters relative to its 
depth at the point where the surface and the geoid coincide.

Thus, the accuracy of satellite  altimeters can be validated  by 
measuring temperature versus depth at various points in the 
global ocean. An illustration well known in oceanographic circles shows a New Jersey high school teacher deploying an Expendable Bathyhermograph (XBT) into a river for the edification of his students; 
an amateur could validate a one-dimensional section by dropping XBTs from a cruise ship. A dozen dropsondes cost only about \$400, far less 
than the cost of the cruise ticket.

Although oceanography from a cruise ship is a pleasant dream,
 professional oceanographers have a better  way. More than thirty 
nations have collaborated to create a network of 4000 drifting profilers, the ``Argo floats". 
 All have a  set of required basic capabilities, but a subset of floats 
carries additional sensors or descend to greater depths (``deep Argo" floats). Three different entities in the U. S., France and Japan supply 
the basic floats.

Each float  is a little robot submarine that spends most of its time 
drifting at a depth of a kilometer. Every ten days, a piston squeezes a mass of compressible mineral oil to slightly change the buoyancy. The  
float sinks  
  to a depth of about 2 kilometers (or as much as 6 kilometers for the  
``deep Argo" floats). It then rises  to the surface, measuring measuring electrical conductivity, temperature and pressure; from these, salinity and density are computed. The float downloads its data to a satellite in 
about twenty minutes and then descends a  thousand meters to 
await ten days for the next profiling cycle to begin.

The ideal  spatial distribution of the Argo floats is non-uniform. A higher
density of floats in polar waters is desirable because the Rossby radius of 
deformation (described earlier), a fundamental spatial scale for 
oceanic and atmospheric eddies,  \emph{decreases} with increasing \emph{latitude}.

On a flat, non-rotating earth, the optimum distribution of Argo floats
would be spatially uniform. However, the whole population of vortices generated by baroclinic instability would be missing from the 
flat-earth ocean.

The Argo floats are but one of a new generation of \emph{in situ} observing systems that check and cross-validate satellite data.

%The latest generation of autonomous observers are called ``gliders". Like Argo floats, gliders change buoyancy by squeezing mineral 
%oil. Unlike Argo floats, which move only up and down, gliders move horizontally for long distances. Gliders

\section{Equatorial and Coastally-Trapped Waves and ENSO}

El   Ni\~{n}o-Southern Oscillation, known by the acronym ENSO, is a global, recurring part of the climate.
  The trade winds exert a wind stress on the oceans that, to balance, requires the sea surface to 
  \emph{tilt} from low at the eastern coastline to high at the western coastline. When the trade winds weaken, 
the thermocline alters from tilted to horizontal. This adjustment does not happen instantaneously throughout 
the whole ocean; rather it propagates eastward as a wave pulse.

This pulse is composed of a mixture of waves, but the mode with the simplest spatial structure,
which is also the one with the largest 
energy, is the ``equatorial Kelvin wave" in the ``first baroclinic mode". The Kelvin wave is concentrated within a narrow 
waveguide that extends from three degrees south latitude 
to three degrees north latitude. This waveguide is generated by refraction, which in turn is created by  the variations of the Coriolis force with latitude. The 
Kelvin wave requires about two months to cross the width of the Pacific Ocean.

When the Kelvin wave  collides  with the Americas, ocean temperature off Central America rises sharply. 
In normal conditions with large tilts, strong upwelling brings up nutrients from below.
  This upwelling supports a huge population of anchovies. These fish were for centuries 
the main export of Peru. When the water turns warm, however, the upwelling stops, the anchovies die
and the Peruvian fishing industry is soon in a very bad way. Peruvian fisherman noticed that these hot water 
episodes tend to occur around Christmas time, so the name ``El Ni\~{n}o", ``the
Christ Child" was attached to this anchovy-killing ocean phenomena.

The mischief is not confined to the equatorial waveguide. After the equatorial Kevin wave collides with the coast of Central America, 
    most of its  energy
radiates north and south as ``coastal Kelvin waves". These are so-named because Coriolis force 
 creates a coastal waveguide similar to the equatorial waveguide. The result of the poleward propagation of coastal Kelvin waves is a strip of hot water extending far south of the equator and as far north as California.

  The evaporation rate is an exponential function of ocean temperature, so huge amounts of water vapor 
form above the hot patch created by coastal Kelvin waves. As this moisture-laden air is pushed into 
the mountains of California and is driven up the western slopes, the rising air is adiabatically cooled, yielding clouds first and then torrential rains. The mansions of movie stars in Malibu slide into the sea when the supporting soil is converted into mud.

Meanwhile, Australia goes through severe droughts, the ``Great Dry". When the equatorial Kelvin wave
  sneezes, the whole world catches pneumonia.

The ocean-induced changes in the atmosphere initiate the next phase of the atmospheric cycle known
as the Southern Oscillation. The period of ENSO varies irregularly from five years to seven years.

The equatorial Kelvin wave is equatorial only because the Coriolis force varies with latitude. The 
coastal Kelvin wave hugs the coast only because Coriolis force balances the pressure force perpendicular 
to the coast. El Ni\~{n}o and the Southern Oscillation would not exist without Coriolis force. ENSO 
cannot exist on a flat earth.

\section{Tornadoes}

Dust devils are small vortices that are visible as helical trajectories of sand or dust or fallen leaves or snow. Dust devils are cyclostrophically-balanced, that is to say, centrifugal force (always outward) balances a pressure force which is inward towards the minimum of  pressure on the axis of the vortex. There is no obvious reason why dust devils should spin one way in preference to another, and indeed cyclonic and anticyclonic dust devils are found in roughly equal numbers.

Tornadoes are also cyclostrophically-balanced, but are almost always cyclonic. Why?

The answer is that tornadoes derive their vorticity from their parent, which is either a hurricane or a medium scale weather system called a ``mesocyclone". Mesocyclones in turn inherit their vorticity from large scale weather systems, so-called ``synoptic [scale] cyclones".

Vorticity is a measure of the rotation of a flow at a given point. Strictly speaking, the vorticity is a \emph{vector}; however, for many kinds of dynamics only the vertical component of the vorticity is relevant. In an abuse of notation, the same label is applied both to the vorticity vector and to the scalar which is its vertical component. Henceforth vorticity will be used to denote the vertical component of the vorticity only.

The vorticity at the surface of a flowing brook or river can be measured by dropping a twig in the water. 
The vorticity is twice the angular frequency of the rotation of the twig, which thus serves as a very low-tech vorticity meter. (The 
factor of two is a mathematical convention of no physical significance.)

In the absence of torques, angular momentum is conserved, but not the vorticity itself. A good example of vorticity change is a spin in figure skating. The skater's first task is to apply torque to make   her body   rotate with the arms flung wide and the skater so low as to be almost sitting on the ice. The angular momentum of a small bit of matter with respect to a given axis of rotation is given by the product of the mass of the small bit of matter multiplied by its rotation rate and the distance from the rotation axis. To spin very fast, the skater stands up and pulls her legs in tight, as close to the spin axis as possible. Simultaneously, the arms are pulled in and thrust over the head to move the mass of the arms as close to the rotation axis as possible.  Because the distance of much of her body has been moved closer to the spin axis, conservation of angular momentum requires that the rotation rate increase. Without applying any additional torque, the skater  spins much faster. Her vorticity is greatly increased when she stretched her body vertically along the rotation axis.

A similar process called "vortex stretching" allows flows to tremendously concentrate and intensify vorticity. The most extreme example is a tornado. 

In addition, the merger of vortices of the same sign combines two small vortices into a single larger vortex,
 a process that also contributes to tornado ferocity.

This transfer of vorticity from large spatial scales to small 
is a general tendency in fluid mechanics. This ``cascade" is stopped 
only when the spatial scale is so small that viscosity becomes important, as in boundary layers.

Lewis Richardson, the Father of Numerical Prediction, wittily 
summarized the vorticity cascade with a paraphrase of 
 Jonathan Swift's two-line poem, ``Big fleas have little fleas that feed and scratch and bite'em, and little fleas have lesser fleas and so \emph{ad infinitum}." Richardson wrote,
\begin{quotation}
``Big whirls have little whirls that feed on  their vorticity, and little whirls have lesser whirls and so on to viscosity."
\end{quotation}

Thus, although Coriolis force is not  significant  in the balance  of a mature tornado, the vortex  ultimately derives its vorticity from the rotation of the earth through
 the vorticity cascade from ``big whirls" to ``little whirls". On a \emph{non-rotating} planet, \emph{tornadoes would not exist}.

\section{Meddies}

As noted earlier,  ``Water Always Seeks Its Own Level" is recited endlessly by flat earth proponents, 
never with any proof except personal observation. This false ``law"  is interpreted by Flat Earth proponents to mean that the ocean surface always relaxes to a state with a horizontal, uncurved surface,

This is not an accurate description of  the Straits of Gibraltar. The Mediterranean Sea is very shallow and very sunny. There is so much evaporation that the  mean sea level of the Mediterranean Sea is a few centimeters  \emph{lower} than that of the Atlantic Ocean. Cold water, of relatively low salinity, pours steadily from the Atlantic Ocean into the Mediterranean Sea as a surface current eastward through the Straits of Gibraltar. Underneath this eastward surface current there is an undercurrent in the opposite direction. Although warm, the Mediterranean bottom water is very salty and therefore very dense. As it falls into the Atlantic, the salty Mediterranean water sinks to about a kilometer below the surface of the Atlantic. The flow breaks up into lenticular\footnote{``Lenticular" means `lens-shaped".} vortices which are called ``Meddies" by oceanographers \cite{Cartonetal10}.

Meddies are always anticyclonic, that is, the spin is clockwise as viewed from above. These vortices are only a thousand meters thick in the vertical direction, but have diameters of 20   kilometers to  100 kilometers!  
  Lifetimes range from five to twenty-five years.

Meddies are easily detected by sounding the ocean with 
an expendable bathythermograph, an inexpensive device [as little as \$35 on eBay!] that simultaneously measures 
pressure and temperature as it falls into the sea, tugging a communications wire behind. Meddies are 
huge lenticular patches of very salty but also very warm water embedded in the cold and less salty
waters of the Atlantic Ocean, and thus are easily identified with the simplest of equipment.

Why do these vortices form? The technical answer is ``conservation of potential vorticity". Although the details are too complicated to be presented here, the important point is that the source of the spin  is the Earth's rotation. This class of vortices would not exist were  it not for the Earth's rotation.

\section{Hemispheric Differences}

Weather and climate are somewhat different between the two hemispheres, but the usual explanation is that these differences are due to the asymmetric distribution of the continents:
there is much more land and much less water in the northern hemisphere than the southern hemisphere. 
In addition, the Arctic is a thin sheet of constantly moving and deforming ice above ocean whereas the Antarctic surface is permanent, long-lived ice heaped kilometers thick above solid ground.
When general circulation models are run on supercomputers with land omitted, so-called ``acquaplanet" simulations, the hemispheric differences \emph{vanish} on the \emph{globe}.

On the aquaplanet flat earth \emph{disk}, there are considerable hemispheric differences. It will seem odd and inaccurate to apply ``hemisphere", which literally means ``half-spherical", to a non-spherical geometry like a disk, but no better terminology is available. The ``disk northern hemisphere" is a disk of radius $R$ (shaded in Fig.~\ref{FigFlatEarth_Sch_Map}); by ``disk southern hemisphere" we mean the  annulus of outer radius $2 R$ and inner radius $R$ (unshaded in Fig.~\ref{FigFlatEarth_Sch_Map}). The area of the northern hemisphere disk is $\pi R^{2}$; the area of the entire flat earth, defined by $r \in [0, 2R]$, is $4 \pi R^{2}$. Subtracting the area of the northern hemisphere from the area of the whole gives
\begin{eqnarray}
\text{Area(disk southern hemisphere)} = 3 \, \pi R^{2}
\end{eqnarray}
Thus, the area of the disk southern hemisphere   is   \emph{three times} the area 
of the disk northern hemisphere.

This will create huge differences in weather and climate between the hemispheres due solely to geometry. In reality, the observed hemispheric differences 
are accurately explained by the  difference    in land/sea fraction and the ice-over-water, ice-over-land difference   between the Arctic and Antarctic.

\section{Summary}

The weather on a non-rotating flat earth would be profoundly different 
from that actually observed. The following elements of our atmosphere  and ocean would 
be missing on a flat earth:
\begin{enumerate}
\item hurricanes and typhoons
\item synoptic cyclones and anticyclones; baroclinic instability
\item Hadley cell descending in the subtropics;
\item trade winds
\item prevailing westerlies
\item equatorial Kelvin waves
\item coastal Kelvin waves
\item El Ni\~{n}o/Southern Oscillation (ENSO) coupled air-ocean cycle
\item tornadoes
\item Rossby waves
\item The Gulf Stream
\item Gulf Stream rings
\item Meddies
\item Sverdrup flow
\item Geostrophically-balanced jets
\item Ekman  layers.
\item Large temperature differences between Costa Rica and Calgary.
\end{enumerate}
 This list is not exhaustive; a good exercise for students is to extend this catalog.

It must be emphasized that the qualitative arguments given here can only be turned 
into a comprehensive, quantitative understanding by applying numerical models and sometimes
 singular perturbation theory. Both the qualitative thinking (here) and supercomputer number-crunching (a topic for future work) are 
essential parts of earthly climate forecasting and exoplanet understanding.

In the appendices, we provide further development of key ideas used above, and also discuss the misconceptions 
of flat earth videos. Because the video creators have [if flat earth evangelists!] no background or training, the mistakes
  are those of introductory physics. The difference is that freshman physics errors are unlearned, corrected and 
forgotten whereas video mistakes are illustrated with gaudy cartoons and quotes that make it easier to understand 
how elementary ideas are misconceived and misinterpreted.

\appendix

\section{Scale Analysis}

\begin{quotation}
``Wheeler's First Moral Principle: Never do a calculation unless you already know the answer."
\end{quotation}
\hspace*{0.5in} --- John A. Wheeler (1911--2008), coiner of the term ``black hole", via 
private communication from Joel R.  Primack (1945--), an undergraduate student of Wheeler's.

\bigskip

Wheeler's First Moral Principle at first glance appears nonsensical. His point is that it is
important to \emph{estimate} the answer before doing a detailed calculation. Estimates 
abort many unfeasible projects in the womb. Estimates catch many misplaced decimal points and 
other errors in projects that are continued through detailed analysis.

Scale analysis is almost universally employed in geophysical fluid dynamics for estimates and qualitative understanding.. The full Navier-Stokes 
equations contain a large number of terms, but for a given species of wave or vortex, most of 
these terms are insignificant and can be neglected. Scale analysis simplifies theory and numerical models,
but it also improvements conceptualization by, as it were, stripping away the dead branches and 
underbrush.

The first step is to estimate an order-of-magnitude for the scales of temporal and spatial variation, velocities, pressure gradient and density fluctuations 
 in the phenomenon. For synoptic scale cyclones, for example, observations over many years show that 
the horizontal velocity scale U is about 10 m/s, the vertical velocity scale $W$ is about a thousand 
times smaller, the horizontal and vertical scales are 1000 km and 10 km, the time scale is about $10^{5}$ seconds, etc. A full discussion is given in Holton and Hakim \cite{HoltonHakim13} and in 
Hoskins and James \cite{HoskinsJames14}. A partial derivative of a variable $u$ with respect to a coordinate (time or space) is the magnitude of $u$ derived by the scale of the coordinate.

Scale analysis shows that for  synoptic cyclones, the vertical momentum equation is dominated 
by the two terms of hydrostatic balance. The largest of the non-hydrostatic terms is roughly 10,000 times smaller. 
 Hydrostatic balance is not a conjecture or hypothesis or theory. It is a \emph{description}.

Similarly, scale analysis of the horizontal momentum equations shows that the dominant terms are 
geostrophic balance: the pressure gradient is balanced by Coriolis force; the remaining terms are an 
order of magnitude smaller. This, too, is not a conjecture but a description.

Scale analysis is not a theory. It is more powerful than a theory.
  It is a description.

In a photo of my wife's grandniece Madeline, her flame-colored curls are not a theory. 
They are a description.

A theory is a union of related descriptions and a hypothesis that make a prediction.

A set of pictures suggests the hypothesis of ``always red  Madeline theory": that she will display 
curly red hair in all her future pictures.

This sweeping hypothesis  is easily falsified: on a snowy day, her hair might 
be completely hidden by the hood of her snow parka. Madeline theory can be protected 
from this failure mode by adding the clause, "if her hair is visible".

The hypothesis of this weaker, less broad Madeline theory is also easily falsified: as a  
teen, she might dye her hair a different color. Madeline theory can be protected 
from this failure mode by adding another ``if clause" and restating the 
Madeline hypothesis as, ``Madeline will show curly red hair in her pictures if her hair is showing 
and if her hair has not been dyed."

But this is still not quite right; when Madeline is middle-aged, her hair may turn to gray or white, 
so we must add ``if age has not turned her hair gray". Madeline has just turned two, so 
this last will not be relevant for decades!

Nicholas Tassib Taleb has popularized the concept of a ``black swan": a rare, extreme outlier that turns 
the financial world upside  down and ruins a multitude of banks and investors. Scientists merely add ``if clauses"  after black swans are discovered: 
``All swans are white" becomes ``All swans are white except in western Australia".

\section{Hand-of-God Physics and Vast Overestimation of Coriolis Effects}

A common conceptual error in the flat earth 
literature generates estimates of the effects 
of Coriolis force that are orders  of magnitude too 
large. The difficulty is one of inadequate care 
in conceptualization of experiments. As such, this and similar mistakes are common in beginning physics and meteorology students. This mistake 
is therefore illuminating even in its very wrongness.

The misconceptualized Coriolis force experiments have in common a moving object that is deflected from a straight path: a bullet, a cannon shell, or a helicopter or aircraft. 
It is useful to examine events through different 
reference frames. One is that of a spacecraft which is motionless with respect to the earth 
and sees the planet spinning about its axis; this is an inertial reference frame, implying that Coriolis force 
is not visible to the observer in this reference frame. The 
other is the noninertial reference frame of an 
observer at a fixed location on the earth near
the gun or takeoff point.

Reference frames are not physical states; they are \emph{windows} into a physical 
state. When we look from the outside to see who is chatting in the parlour, we obtain a 
different view and can collect different information by looking through the small window
 on the side of the house versus  the big bay window on the front of the house. 
  The conversation in  the parlour, the physical state,  is unchanged by our choice of windows (frames of reference).

 To make the example 
 concrete, 
assume the departure point is on the equator (0 degrees latitude) on the Greenwich meridian (0 degrees longitude).

At takeoff time, the helicopter/bullet/shell 
and the earth are both motionless in the earth-fixed reference frame (columns five and 
six in Table~\ref{TabCo}) and both  are  traveling 
eastward at 1670 kph in the spaceship's non-rotating reference frame (columns two and 
three in Table~\ref{TabCo}). 

In midflight, 
  flat earth proponents argue that  ``Aircraft in flight are independent of the surface once they take off ... The helicopter and surface
are moving along as one but the moment
the helicopter takes off vertically ... it becomes free and
independent of the surface of the earth. It is now
hovering in the air and the earth will
carry on, leaving the helicopter behind." \cite{PhuketWordCoriolisHelo}. In other words, for most 
of the flight, the helicopter  is hovering, motionless, in the inertial, non-rotating spaceship reference frame while the surface earth 
whizzes at 1670 kph underneath the hovering helicopter. At the end of a twelve-hour flight,
the helicopter will be above the international 
dateline, having traveled more than 20,000 km westward 
merely by hovering up and down. This prediction is a gross misinterpretation 
of the standard physics of a rotating globe. Flat earth proponents agree this scenario is 
absurd but draw the erroneous  conclusion  that  the earth cannot be a spinning 
globe! 
    
Equivlently, as ``Phuket Word"  puts it in \cite{PhuketWordCoriolisHelo}, ``The atmosphere does not stick to the earth and the earth does not move beneath aircraft - a simple proof that the earth is flat and stationary."

The trouble with this analysis is that the 
helicopter, which initially has no horizontal 
 motion in the \emph{ground observer's} reference 
frame, somehow becomes motionless in the 
\emph{spaceship observer's} inertial, non-rotating
    reference frame. But this requires altering 
the helicopter's velocity in the inertial 
frame from 1670 kph  to 0! What can provide 
such a huge velocity change?

The soccer superstar Diego Maradona scored 
the World Cup-winning goal in 1985 by quite illegally punching
   the ball past the goalkeeper with his hand.
   Unwilling to admit that he had cheated,
 Maradona said afterwards, that the goal was scored, ``A little with his head, and a little with the hand of God".

A punch by the Hand-of-God seems as good a hypothesis as any
for the force that produces this huge and mysterious change of the horizontal velocity of 
the helicopter in the flat earth misconception of physics.

 There are further difficulties. If the wind remained at rest at all heights in the rotating/ground-observer reference frame, 
then the helicopter, hovering at rest in the inertial frame, would be buffeted by a wind moving
past the helicopter eastward at 1670 kph.  Consequently, this scenario requires an additional 
mechanism which we shall dub the ``Albertus-Magnus-Superstrong-Steady-Wind-Force" which somehow maintains,  in the ground-observer frame, a vertical wind profile of the longitudinal velocity which is zero at the ground and -1670 kph at the altitude of the hovering helicopter, more than twice the speed  of sound. (The red digits in Table~\ref{TabCo} 
show the required wind speed; It is not necessary for the mean wind to change with time; a steady wind varying strongly with height is sufficient if the helicopter chooses its rate of ascent or descent to match  its time-dependent horizontal speed to that of the wind.)

The great Dominican 
theologian and experimental natural philosopher, St. Albertus Magnus, remarked in \emph{De Mineralibus} (Book II, Tractate ii, Ch. 1), ``For it is [the task] of natural science not simply to accept what we are told but to inquire into the causes of natural things." Although the patron saint of scientists, the good bishop would be hard pressed 
to inquire into the ``causes" of the   Albertus-Magnus-Superstrong-Steady-Wind-Force because it seems 
unrecognized in Hand-of-God physics that such a wind shear-maintaining force is even required.

 %%%%%%%%%%%%%%%%%%% TABLE HAND-OF-GOD  %%%%
\begin{table} \caption{\label{TabCo} Coriolis force experiments: Velocities $v$ and longitude $\lambda$ for helicopter [``helo"], ground and atmosphere at 1 km altitude}
\begin{tabular}{|c|c|c|c|c|c|c|c|} \hline
   & \multicolumn{3}{|c|}{spaceship (inertial frame)} & \multicolumn{3}{|c|}{ground observer (rotating frame)} & \\
 time & helo $v$& earth $v$ & air $v$ &  
helo $v$& earth $v$& air $v$ & helo $\lambda$ \\
   \hline 
$t=0$ & 1670 & 1670 & \textcolor{red}{0} 
  & 0 & 0      & \textcolor{red}{-1670} & $0^{o}$
\\  
$t=t_{1}$  & \multicolumn{7}{|c|} {Hand-of-God changes velocity in inertial frame from 1670 kph to zero}    \\ 
$t=6$ & 0 & 1670  & \textcolor{red}{0} &
   -1670  & 0  & \textcolor{red}{-1670} & $- 90^{o}$
\\ 
$t=t_{2}$  & \multicolumn{7}{|c|} {Hand-of-God changes velocity in inertial frame from 0 kph to 1670 kph}      \\ 
$t=12$ & 1670 & 1670  & \textcolor{red}{0} &
   0  & 0  & \textcolor{red}{-1670} & $- 180^{o}$
\\  \hline
\end{tabular} \end{table}
% ***************

Lest one suppose that this misconception was 
solely that of the video maker ``Phuket Word" from 
 which came the quote above, we shall offer 
additional instances.

For example, Paul Linberg alias ``Paul on the Plane" in 
\cite{PaulonthePlaneCoriolis} analyzes a rifle bullet fired vertically. Without the Hand-of-God, 
classical mechanics predicts that the bullet will land on the top of the rifle that fired it, neglecting 
crosswinds. (The Coriolis force does produce a small horizontal deflection on the way up, but this is canceled by Coriolis deflection in the 
opposite direction on the way down.) 

\begin{figure}[h]
\centerline{\includegraphics[scale=0.5]{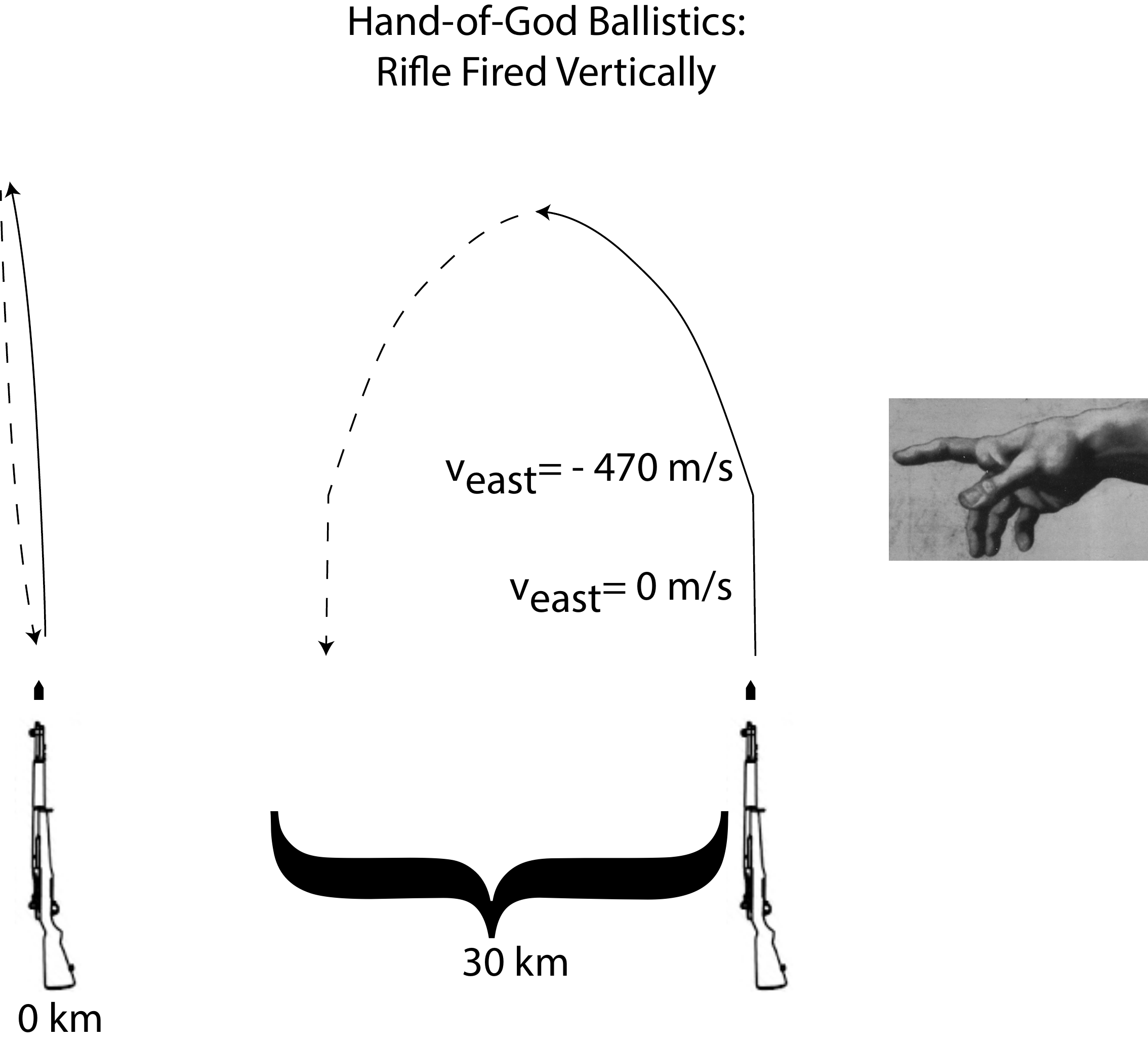}}
\caption{When a rifle is shot straight up in the air, the bullet is deflected a few centimeters westward 
by Coriolis force on its upward flight, but it is deflected eastward on its downward flight so that it lands 
 on the rifle as shown on the left. In Hand-of-God physics (right), the bullet is given a westward velocity of near half a kilometer per second (as seen by an observer on the ground) by Divine Intervention  so that the bullet has no horizontal 
velocity in the inertial non-rotating reference frame. Experimentally, the bullet is in the air for a couple of 
minutes. Hand-of-God physics then predicts that the bullet will land thirty kilometers west of the rifle. The true situation 
on a rotating spphere is shown  on the left: without intervention by the Hand-of-God, 
the bullet will land very close to the rifle with the distance from the gun depending primarily  on the strength of the horizontal breeze.}
\label{FigVertical_Rifle_Hand_of_God}
% Drawn by me in Illustrator as Vertical_Rifle_Hand_of_God
\end{figure}

At minute 35 of the video, Paul asserts 
that for the rotating globe, \begin{quotation}``Watch this demonstration of a bullet fired straight up in the air. Its flight time is about two
minutes.  It lands almost precisely where it was shot from [as predicted by standard physics without the Hand-of-God].  If the Coriolis Effect was a real thing  and
the earth was spinning, this bullet should have landed about 21 miles away." \end{quotation} 
His prediction and what is actually forecast by physics are shown in Fig.~\ref{FigVertical_Rifle_Hand_of_God}.
%(Three minutes later in his video, he begins a detailed 
% debunking of Newton's theory of gravitation.)

 % **********************  FIGURE  ***********
% ***************************************

The prominent group of debunkers known collectively as the ``Ghostbusters" show a 
video by Ron Hagberg  
in \cite{GlobebustersCoriolis}. Hagberg examines a battleship firing at its maximum range (25 miles) due south from Singapore (which is a negligible one degree from the equator). Coriolis force would 
deflect the shell more than 20 miles eastward so
that it would have to travel 33 miles --- 8 miles beyond its official maximum range --- to hit the target. R. Knodel, Globebuster and host, says
  ``I couldn't agree more. ... Ridiculous that globers claim the air is in lockstep with the ground!" \cite{GlobebustersCoriolis}.
 
In reality,  the battleship shell  fired at an elevation of forty-five degrees would be deflected by Coriolis force by less than one-tenth of a mile, not 
twenty-one miles unless the Hand-of-God is applied 
to change the velocity of the shell in the inertial reference frame by 1670 kph soon after 
firing, and  then the Hand-of-God strikes again  in the opposite 
direction just before the shell impacts.

Nathan Thompson, who has no occupation except 
flat earth evangelism and charges \$500 for a video debate, asserts in \cite{ThompsonNathanFluidsDec2019}, 
"He [Felix Baumgartner]  ascended for three hours [and then jumped from his balloon at 128,000 feet of altitude] and landed
back in New Mexico; he should have landed
somewhere in the Pacific Ocean." Thompson assumes that shortly after takeoff, Baumgartner acquires a 
horizontal longitudinal velocity of -1670 kph so that the earth would have rotated  underneath him by one-eighth of the length of the circle of latitude while his balloon was stationary in the inertial reference frame. 
Similarly, some force restores the parachutist to zero horizontal motion in the 
ground observer's rotation frame at some unspecified altitude before he lands. 
Thompson's velocities for the parachutist are correct only if the Hand-of-God acted once on the ascent 
and once on the descent.

However, in the same video, he writes on one of his slides
 that for an airplane at the equator, ``North-South flights would face a crosswind 
of 1000 mph" [1670 kph]. In an animated slide, he shows people [at rest relative to the rotating 
earth] being swept away by the wind. Is the velocity zero at surface, as apparently true for 
Baumgartner's launch and landing, or ferociously westward at ground level?\footnote{The video contains 
several additional errors. Thompson concludes that solid body rotation of the atmosphere is impossible because ``What natural force could do this?"; the correct answer is ``viscous shearing stresses".  He adds, ``This is
an indisputable law of fluid dynamics: If
the earth were spinning, the atmosphere
would move separately from the earth and
I'm of the opinion it would blend the
atmosphere into one homogeneous gas". This is nonsense; the spinning globe is not the blade of 
a mixmaster; he gives no motive for his ``opinion".  
While showing water droplets flung away from  a rotating tennis ball, he says, 
``What would the oceans do on a spinning earth? 
Well they'd fly off because any ball
with water on it that you spin --- all
that stuff flies off";  his oceans are apparently 
 liquid Cavorite  which annuls gravity in H. G. Wells' novel \cite{Wells01}. The flat earth literature and videos usually contain multiple errors besides an incorrect shape for the planet; one must ignore some inconsistencies in order to stretch 
the mind by trying to work out the geophysics of worlds that never were.}
 Thompson is not consistent even within a given video.

\begin{figure}[h]
\centerline{\includegraphics[scale=0.7]{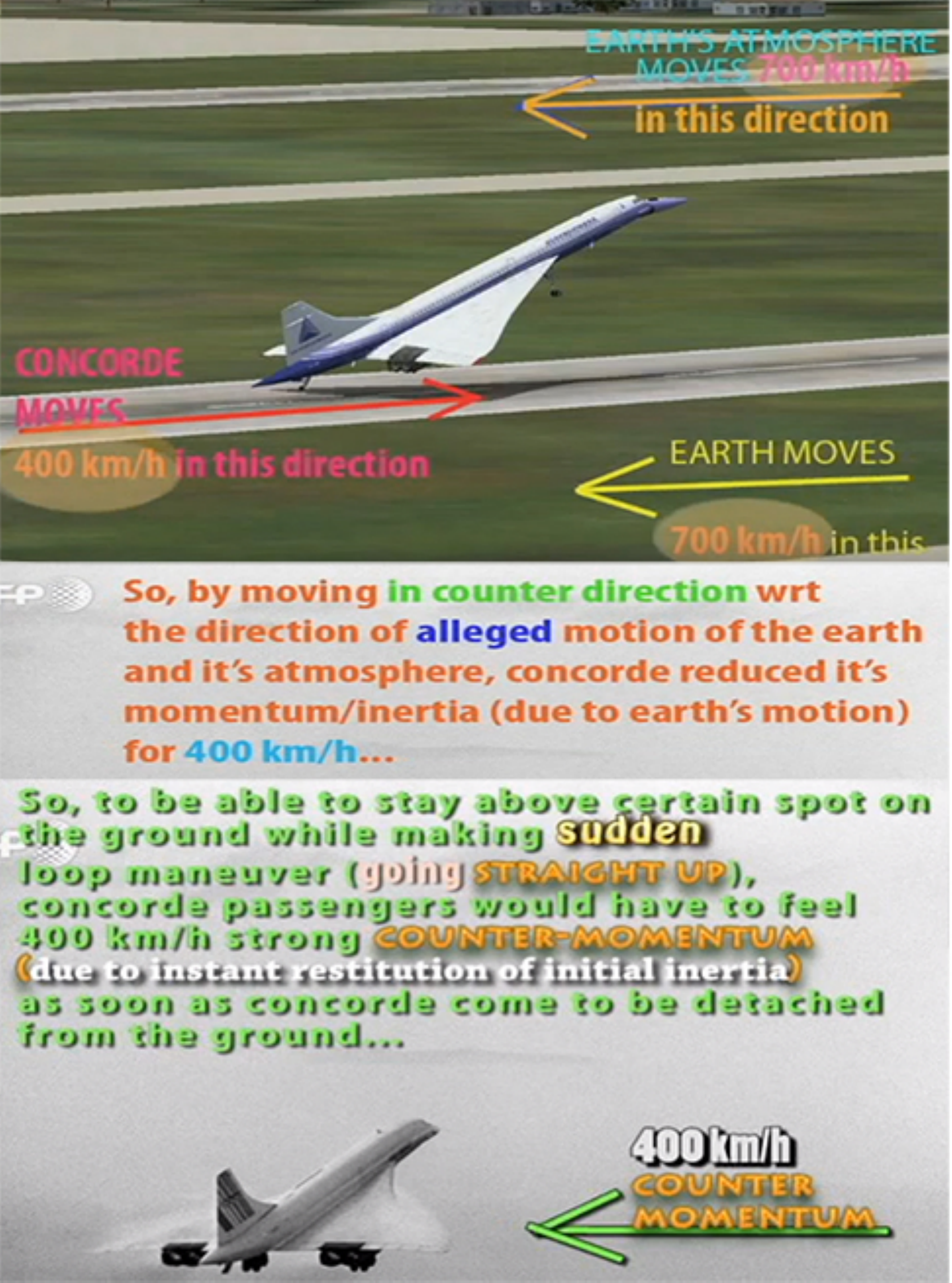}}
\caption{At a latitude where the linear velocity of the spinning earth is 700 kph, 
the Concorde takes off. By flying at its maximum velocity through the air of about 
1100 kph, the  Concorde can manage to fly 400 km eastward. The videomaker argues that 
during the ``instant restitution of initial inertia as soon as Concorde come to be detached 
from the ground", the passengers would receive  titanic accelerations.
 No such passenger-and-crew squishing accelerations are observed; therefore the earth is flat. This is true only if one 
enlarges standard physics by adding, for ``instant restitution",  the Hand-of-God.}
\label{FigConcorde_Thought_Experiment_odiupicku}
% odiupicku channel, assembled from three frames of Thought experiment ({Concorde} argument
\end{figure}

Another flat earth advocate is ``odiupicku". Fig.~\ref{FigConcorde_Thought_Experiment_odiupicku}
 combines three screenshots of a video \cite{odiupickuConcorde}. The author falsely predicts 
that standard physics on a rotating sphere requires that the passengers and crew be crushed by 
enormous accelerations at the moment of takeoff; this ``instant restitution of initial inertia as soon as Concorde come to be detached 
from the ground," exists only 
 in his misinterpretation of physics.   ``Instant restitution of initial inertia" is possible 
only if the Concorde were struck by the Hand-of-God. ``Instant" velocity changes 
are possible only in bad science fiction novels written by innumerate college dropouts.

 % **********************  FIGURE  ***********
% **************************************

Another well-known video maker, Quantum Eraser [John Stunja], said  in \cite{ThompsonWaveCoriolis},
\begin{quotation}
``It's very simple you need two separate
reference frames. You need the non-inertial reference frame that's the
spinning ball earth and an inertial
reference frame; that's the atmosphere.

%You need those two. You can't do it
%without those two. it's only those two. So
%it doesn't matter right if the earth is
%spinning and a plane takes off anywhere
%in China and Japan and Los Angeles so they
%have straight flights, one-way flights
% right to LA. 
%
If the earth is
spinning in your face which it has to be
because you leave the non inertial
reference frame when you take off.   
Commercial
flights fly at about 500 miles an hour,
so you have to add those speeds, though,
500 miles an hour for the plane speed
plus the rotation speed that's coming in
your face. 
    That equals 1360 miles
an hour. That means that that flight
should take 1.5 hours. It's just 
what it is, right? So how many Charlotte,
North Carolina flights to LA take one
and a half hours? 
None. They're all four and a half
hours.   It's over [for the globe earth model]. It's, it's done! There you
go!"
\end{quotation}

All  of these videomakers seem confused by frames of reference. In remarks not 
quoted, it appears that all understand that there are two useful frames of reference:
  the inertial reference frame of the spaceship observer and the non-inertial, steadily rotating reference
frame of a ground-based observer. The mistake is identifying \emph{frames of references} with 
\emph{physical states}. This misconception in turn leads to interpreting a change of frame 
as equivalent to an acceleration, despite the absence of a force other than the Hand-of-God!

The misconceptions of flat earth scientists are likely similar to those of students with no 
previous experience with physics, and thus of interest to teachers of introductory physics.

Why the mistake? Partly, it is a failure of 
damage control. Science is like wrestling alligators; the problem squirms and wiggles, always trying to escape, always trying to bite. Good scientists are active sceptics, always on the lookout for 
bias, approximation and experimental errors, an excess of free parameters. and incomplete theories.

  \emph{Checking} is as important as \emph{derivation} and 
\emph{calculation}. Veteran scientists and 
engineers, like middle-aged gladiators, bear 
the scars of their errors.

Unfortunately, flat earth science is outsider 
science, and outsider scientists are amateurs,
  untrained in error-catching. A table like 
our Table~\ref{TabCo}, which lays out all the 
stages from all relevant reference frames, is very 
helpful in indentifying inconsistencies. 

A related peril is what fighter pilots call 
``target fixation". Manfred Freiherr von 
Richtofen, the ``Red Baron", the highest-scoring 
ace of World War I, was pursuing his intended eighty-first victim in April, 1918, when the rookie
 pilot, Wilfrid May, somehow survived von Richtofen's  first burst and 
initiated a long dogfight. The Red Baron lost track of the other Canadian planes, and squadron leader Brown was able to fire a machine gun burst from long range.
  May's fighter lost height with each turn, but
the Red Baron ignored his own maxim, ``stay high!", repeated endlessly to his fliers, and chased his target downward until  
 Australian infantry opened up on him. Finally, a bullet hit him in
the side and drilled both lungs. The 
 source of the fatal bullet is controversial even  a century later because in his fixation on May,
 the Red Baron exposed himself to a whole host of enemies.

Flat earth proponents who succumb to ``Hand-of-God" physics are fixated on a single
mental image: the helicopter/bullet/shell which 
is  horizontally motionless in the inertial 
reference frame of  the spaceship while the 
planet whizzes madly by at 1670 kph. But what 
is the initial state? The final state? How does
 the initial state evolve to the 
motionless helicopter and the rotating planet? How does
one evolve from that intermediate state  to 
the final state in which helicopter and 
planet are once again of equal velocity? Fixating 
on one state in one reference frame when there at least 
three relevant states and two  different but 
informative reference frames is a prescription, metaphorically speaking, 
for being shot down.

A final peril is what one might call ``Phrasebook Physics". Physics and mathematics 
are languages. Thorough study is required to reach fluency in physics no less than 
French. 

Every year, hordes of tourists explore Paris without first learning the language, 
  armed only with a phrasebook containing translations of sentences like, "Which way to 
the toilets? " and ``point me to the Louvre". Unfortunately, many makers of flatearth 
video understand the languages of physics and mathematics at an equally superficial 
level. The phrasebook entries are replaced by Wikipedia defnitions and 
the occasional flashy image.

 Tourists usually are aware of their lack of fluency. 
Flat earth evangelists  are proud of their lack of formal training, which shields 
them from ``indoctrination" with science,  which is merely ``pseudoscience" 
and ``religion".\footnote{Outsider art is similar; the sociologist Gary Alan Fine records 
a woman who argued that she be allowed to exhibit in a show of untrained artists, 
despite the mortal handical of a graduate degree [in law], because she was a 
diagnosed schizophrenic \cite{Fine04}.}

\section{Gravity}

Flat-earthers almost universally believe that the Newtonian and Einsteinian theories 
of gravity are totally wrong. Instead, gravity [they claim] can be replaced by an effect that depends only 
on density and buoyancy where the latter is defined as the relative buoyancy of neighboring 
masses. 
  Since no one in the flat earth community has a working knowledge of differential geometry, 
metric tensors, etc., the sneers at ``bendy space-time" are as silly as a violently negative
reviews of the Great Bulgarian novel, available only in its native tongue, by a literary critic
who can't read a word of Bulgarian. But expertise is hated, and anything requiring a college course 
is pretentious, pompous, dogma or a secular religion and self-evidently wrong.

I have not been able to identify a coherent theory for the density-buoyancy pseudo-gravity. A serious  conceptual difficulty is that gravity is  a \emph{force}
and therefore a \emph{vector}. Density and buoyancy are \emph{scalars}. There is nothing 
in the density-buoyancy theories to explain why the motion is downward; no mathematical 
expression for the force.

Some flat-earthers have creatively added an additional property called ``droppity" (!). But 
we are already far into the \emph{Alice in Wonderland} unreality where words mean ``just what I want them to mean"
instead of the standard dictionary definitions, and magical forces appear wherever needed.

Recently, some flat earth pundits endorsed the ``Electric Universe" theory. This has more magic forces and properties than the wizards of the game \emph{Dungeons and Dragons}, so we shall 
describe flat earth gravity no more.

Tort \cite{Tort14} and Kuzii and Rovenchak \cite{KuziiRovenchak19} 
have calculated the Newton gravity generated by a flat disk 
as a useful homework for physics students. Both papers note 
that their results  imply   that a stable flat earth is impossible.

\section{The Sign-of-Vortices Problem}

When viewed from above (viewed from orbit), hurricanes have fluid rotating counterclockwise around
the minimum of  pressure in the northern hemisphere, but spin clockwise in the southern hemisphere.
  To a first approximation, hurricanes are in geostrophic balance with the inward pressure force 
balanced by Coriolis force. Because the Coriolis parameter is of opposite sign in different hemispheres,
the tangential velocity must reverse sign, too.

Midlatitude synoptic scale vortices similarly rotate with opposite sign in different hemispheres.

How does this work on a diskworld without rotation?  The ``Globebusters" flat earth group has an answer. First, replace gravity by electromagnetic forces.\footnote{They wholeheartedly and uncritically endorse the physics heresy of the ``Electric Universe" described in \cite{ThornhillTalbott07}.} 
  Next,  assume that the diamagnetic interaction of air with electro-gravity produces a strong  
westerly jet centered at the equator which can be identified with the trade winds. They claim that shear vorticity 
on the flanks of the electro-trade winds provides vortices of opposite sign in opposite hemispheres. 

This theory appears derived from the (2015) video by`` FlatEarth RethinkingScience" \cite{FlatEarthRethinkingScience}. This in turn was probably inspired by the 
(2014) video \cite{PhysicsGirlCrazyPoolVortex} by Dianna Cowern on the Physics Girl channel. She created vortices in a swimming pool by pushing a dinner plate through the water. This video went viral with 
over 7.88 million views and inspired ``Physics Girl", who has an undergraduate physics degree 
from MIT, to make three   sequels  \cite{PhysicsGirlFunwithVortexRingsinthePool,PhysicsGirlHowtoMakeaSQUAREVortexRingft3blue1brown,PhysicsGirlHowtoMakeVORTEXRINGSinaPool}.

This dinner-plate-through-the-water analogy was endorsed by the prominent flat earth evangelists Nathan Thompson, JM Truth (Joshua M. Gennari) and Quantum Eraser (John Stunja). 
 In an online discussion \cite{ThompsonWaveCoriolis}, they preached the ``hurricanes as sun wake" theory
 Unfortunately, their illustration  Fig.~\ref{FigThompson_N_Coriolis_Is_Wave_Vortex_of_Sun} may be justly 
renamed the ``Figure of Six Errors":

 % **********************  FIGURE  ***********
\begin{figure}[h]
\centerline{\includegraphics[scale=0.5]{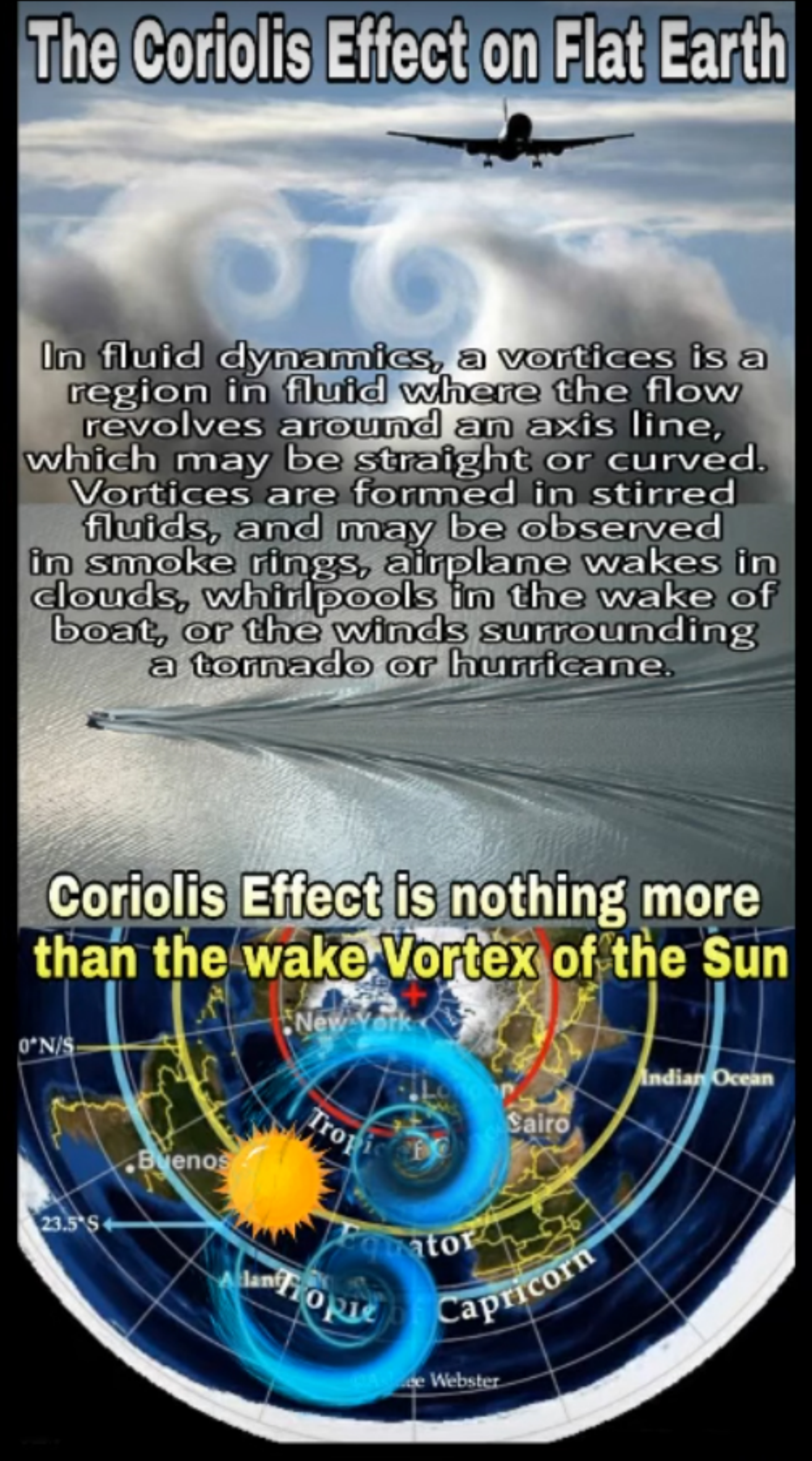}}
\caption{The ``Six Error" illustration: a slide from a Nathan  Thompson video, claiming that  the Coriolis force is the generation 
  of hurricanes and synoptic vortices. Thompson borrowed this from the video ``Flat Earth-CORIOLIS EFFECT The `WAKE' of the sun" by 
FlatEarth RethinkingScience.}
\label{FigThompson_N_Coriolis_Is_Wave_Vortex_of_Sun}
% AFTERSHOW | Quantum Eraser \& JM Truth Join To Discuss Mctoon \& The Plane Truth Debate
\end{figure}  
% ***************************************

\begin{figure}[h]
\centerline{\includegraphics[scale=0.5]{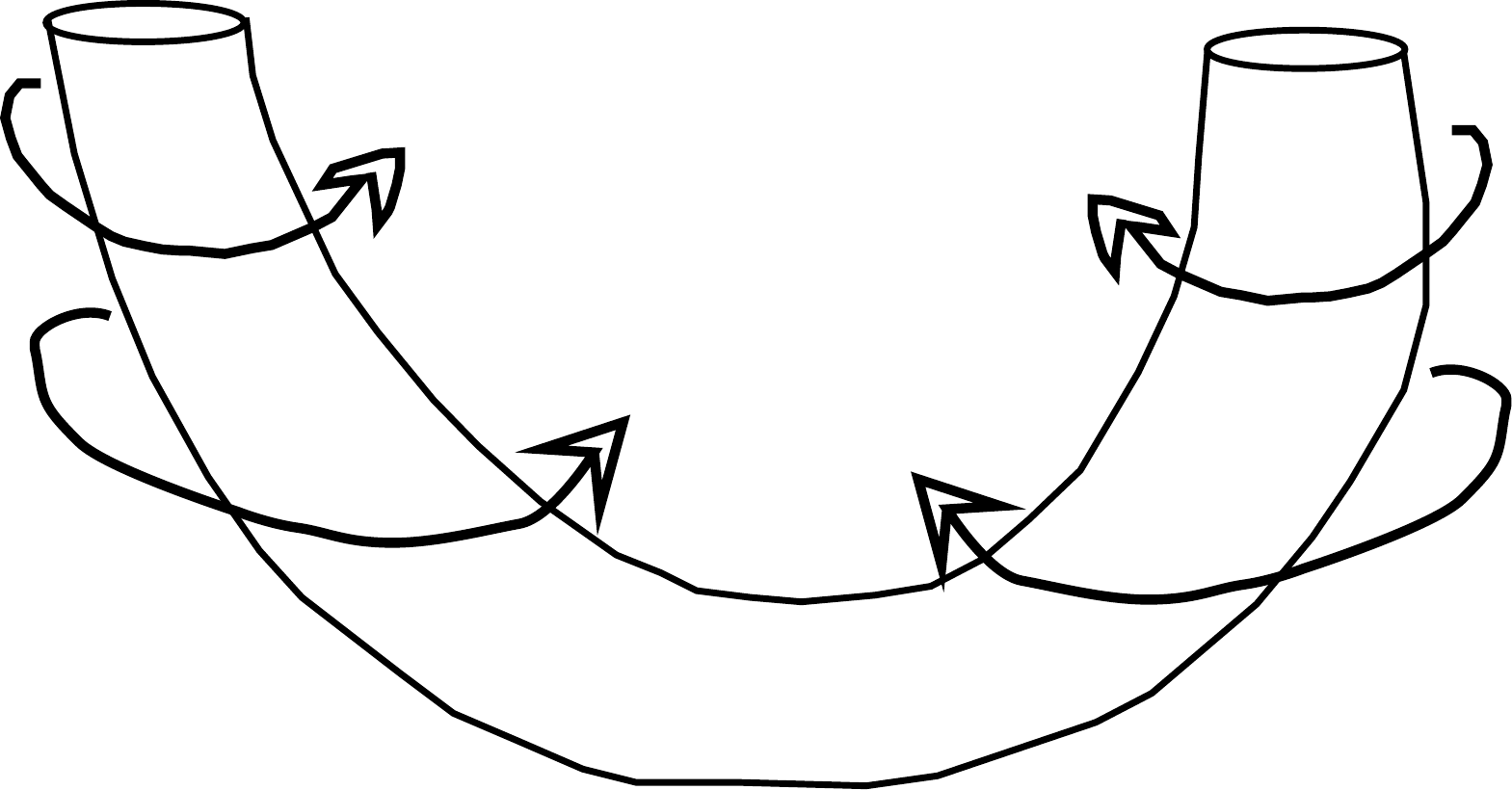}}
\caption{Schematic of a half-vortex ring.}
\label{FigHalfVortexRing}
% Drawn by Boyd in Illustrator
\end{figure}

\begin{enumerate}
\item The wake of a sharply-pointed object,  such as the boat in the grayscale image in the middle of 
Fig.~\ref{FigThompson_N_Coriolis_Is_Wave_Vortex_of_Sun},  is composed of \emph{waves} rather 
than vortices. Lord Kelvin invented the method of stationary phase for the asymptotic evaluation 
of integrals in order to derive a good approximation to this wave-wake, which is now called  the 
``Kelvin ship wake" in his honor.

\item A blunt object like the dinner plate used by Physics Girl, moved to the left (westward), will generate a wake mostly of vortices. However, the vortices are all \emph{anticyclones}, spinning clockwise in the northern 
hemisphere and counterclockwise south of the equator, contrary to the assertion in {FlatEarthRethinkingScience}.

\item Hurricanes are \emph{cyclones}, and therefore the proposal that they are the wakes of a westward-moving 
sun is impossible.

\item In Dianna Cowern's  experiments, her dinner plate was  small in size compared to the pool depth. 
What was generated was a \emph{half vortex ring} as shown in Fig.~\ref{FigHalfVortexRing}. The surface vortices are connected by a tube of vorticity oriented parallel to the surface of the planet, what meteorologists call a ``roll circulation". Observed hurricanes in the northern hemisphere are manifestly not connected 
to a contrarotating hurricane in the southern hemisphere with an immensely powerful roll circulation joining them into a half vortex ring. Hurricanes are 
observed to be  lone warriors, not conjoined twins.

\item There is not a shred of evidence, either observational or calculational,  that the sun can magnetically  push the atmosphere.

\item The statement  that   ``Coriolis Effect is nothing more than the wake Vortex of the Sun" is wrong. 
The Coriolis force is the apparent force that appears in a rotating reference frame; it is not the wake 
of anything.
\end{enumerate}

 % **********************  FIGURE  ***********
% ***************************************

The Ghostbusters/Thompson/JM Truth/Quantum Eraser presentations are devoid of computations, order-of-magnitude estimates and anything 
resembling a quantative description. It is easy to remedy this deficiency.

The mass of the atmosphere is about $5 \times 10^{18}$ kg. The kinetic energy of a flow is the 
product of mass with the square of the velocity. One kilogram accelerated to a velocity of one meter/second 
is a joule of energy.  The dissipation time scale for synoptic scale vortices is about four days;
hurricanes decay perhaps twice as fast. If the jet is assumed to move one-fifth of the mass of the atmosphere at a 
velocity of 10 m/s, the electro-gravity field must supply $10^{20}$ joules every four days, or about
$10^{22}$ joules/year. A terawatt-hour is $8.64 \times 10^{16}$ joules. The total annual global energy consumption 
of human civilization in 2019 is about $150,000$ terawatt hours or about $1.3 \times 10^{22}$, or slightly 
larger than the energy consumption of our idealized atmospheric  jet. 

The yield of the atomic bomb dropped on Hiroshima is estimated at about 
$6.3 \times 10^{13}$ joules. Thus, maintenance of this equatorial jet requires the expenditure annually of the energy of about 160 million Hiroshima atom bombs. The Electromagnetic Pulses (EMP) associated with such
stupendous energies would surely fry every cell phone and every microwave oven on the planet 
on a daily basis.

Furthermore, were such tremendous electromagnetic forces  operating in the natural world, Michael Faraday would surely have discovered 
them a hundred and fifty years ago.

However, even if one is prepared to accept ``electro-gravity", there are still major difficulties. 
The jet is a parallel flow with two bands of opposite-signed \emph{vorticity}, but independent of longitude and
therefore \emph{without} \emph{vortices}. One could further postulate that Kelvin-Helmholtz instability could 
generate vortices, but this is a barotropic instability (drawing on the kinetic energy of the 
jet) whereas observations of atmospheric energetics shows that the instability must be 
baroclinic, drawing energy from the mean  potential energy). Furthermore, there is no
obvious mechanism for generating either hurricanes (which grow over warm water near the 
equator) and synoptic-scale vortices (which have no amplitude in the tropics). Because 
Coriolis forces are excluded, the electro-gravity/jet theory furnishes nothing to balance the 
pressure gradients in large-scale vortices except centrifugal force, which is  a factor of ten 
too small for synoptic-scale vortices.

The electro-gravity ``theory" is really a collection of qualitative assumptions about unobserved 
forces and flows. Words like ``diamagnetic" are tossed around with a meaning far different from the 
standard definition, fairy-forces relabeled to appear  scientific. Even if one is prepared to swallow the handwaving and lack of even the 
crudest numerical estimates, it is impossible to reconcile a world without Coriolis forces
with the weather of the world we live on.

\section{Author's Biography}

John P. Boyd  (jpboyd@umich.edu)
is professor emeritus of atmospheric and oceanic science at the University of Michigan. He taught atmospheric and oceanic dynamics for 42 years. He has published four scientific books \cite{Boyd99z,Boyd98c,BoydBook3,BoydBook4} with two more forthcoming \cite{BoydClenshawBook,BoydImbricateBook}.   He has also published twenty science fiction stories and two hundred and fifty academic journal articles in geophysical fluid dynamics and applied mathematics.
 He was elected a Felllow of the Society for Industrial and Applied Mathematics in 2020.

%John P. Boyd, Dept. of Climate and Space Sciences and Engineering, University of Michigan, Ann 
% Arbor MI 48109-2143.

 % **********************  FIGURE  ***********
%\begin{figure}[b]
%\centerline{\includegraphics[scale=0.5]{.eps}}
%\caption{}
%\label{Fig}
%
%\end{figure}  
% ***************************************

%\centerline{\includegraphics[scale=0.5]{:BezierBernsteinfolder:BernsteinBasisN10.eps}}

%\begin{definition}[COVARIANT and CONTRAVARIANT: BASIS VECTORS]\hfill\\ 
%\label{}
%\end{definition} 

% TABLE TEMPLATE
%\begin{table} \caption{\label{Tab1-1} Title}
%\begin{tabular}{|c|c|c|c|} \hline
%Error  &  polynomial degree $M$  & No. subdivisions     & Cost: $10 Q M^{3}$  \\ \ 
%\end{tabular} \end{table}
% ***************

% *********** TABLE CODE LISTING *********************************
%\begin{table}[h] \caption{\label{Tab} Matlab code  }
%% {\footnotesize This is how to add a note to table.}
%%  \vspace{5pt}
%\begin{center} \begin{tabular}{c}
%
%\begin{lstlisting}
%function fpoly=ChebsumClenshaw(a,x);
%fpoly= (1/2) *(b0 - b3 )  + 0.5*a(1); 
%\end{lstlisting}
% \end{tabular} \end{center} \vspace{5pt} \end{table}  % \addtocounter{tbl}{1}
% ******************************************************

%\nocite{Seife00}

%  \bibliographystyle{siam}  % Specifies references will be
          % written
          %bibtex forthis particular text file using the
          % file
          % siam.bst for the style of the references.

%  \bibliography{Flat_Earth}  % argument of bibliography must be
       %.bib,
       % which is not included in the argument

\end{document}